\documentclass[preprintnumbers,nofootinbib,noshowpacs,eqsecnum,prd]{revtex4}
\pdfoutput=1
\usepackage{graphicx}
\usepackage{amsmath,amsthm,amssymb,amsfonts}
\usepackage{mathrsfs,bbm}
\usepackage{color,hyperref,array}

\definecolor{lgray}{gray}{0.9} 		
\graphicspath{ {figures/} }

\makeatletter       

\renewcommand{\p@subsection}{}

\makeatother

\newcommand*{\eweakgroup}{\mbox{$SU(2)_L \times U(1)_Y$} }

\newcommand*{\unitmatrix}{\mathbbm{1}}

\newcommand*{\twomat}[1]{\underline{#1}}             
\newcommand*{\tvec}[1]{\boldsymbol{#1}}              
\newcommand*{\trans}{\mathrm{T}}                     
\newcommand*{\by}{\!\times\!}                        
\newcommand*{\im}[1]{\text{Im} {#1}}                        
\newcommand*{\re}[1]{\text{Re} {#1}}

\DeclareMathOperator{\tr}{tr}

\DeclareMathOperator{\diag}{diag}		
\DeclareMathOperator{\CP}{CP}			
\DeclareMathOperator{\CPs}{CP_s}		
\DeclareMathOperator{\CPg}{CP_g}		
\DeclareMathOperator{\Pp}{P}		
\DeclareMathOperator{\Ps}{P_s}		
\DeclareMathOperator{\Pg}{P_g}		

\begin{document}

\title{Generalised P and CP transformations in the 3-Higgs-doublet model}

\author{M. Maniatis}
    \email[E-mail: ]{Maniatis8@gmail.com}
\affiliation{Departamento de Ciencias B\'a{}sicas, 
UBB, Casilla 447, Chill\'a{}n, Chile.}
\author{O. Nachtmann}
    \email[E-mail: ]{O.Nachtmann@thphys.uni-heidelberg.de}
\affiliation{Institut f\"ur Theoretische Physik, Philosophenweg 16, 69120
Heidelberg, Germany}


\begin{abstract}
We study generalised P and CP transformations in the three-Higgs-doublet model (3HDM) with Higgs and gauge fields only. 
We find that there are two equivalence classes, with respect to flavour transformations, of generalised P transformations and
there is only one class of CP transformations. We discuss the conditions the potential has to satisfy in order
to be invariant under these transformations. We apply the method of bilinears which we briefly review.
We discuss the relation to the conventional basis, where the potential is written 
in terms of scalar products of the doublet fields. In particular we reproduce the known result that a potential is invariant under
CP transformations if and only if there is a conventional basis where all parameters are real.
Eventually we study standard P and CP transformations in the $n$-Higgs-doublet model (nHDM).
We show that for the bilinears of the nHDM the standard CP transformation corresponds to a diagonal
linear transformation with only $\pm 1$ as diagonal elements. We give this matrix explicitly for arbitrary~$n$. 
\end{abstract}


\maketitle

\section{Introduction}
\label{intro}

One  motivation, decades ago, to study models with an extension
of the number of Higgs-boson doublets was to investigate possible sources of
$\CP$ violation. In \cite{Lee:1973iz} it was shown that in a model with more than
one Higgs field one can have spontaneous CP violation.
In \cite{Kobayashi:1973fv} not only the famous Cabibbo-Kobayashi-Maskawa (CKM)
matrix was introduced, governing CP violation in the standard model (SM) of particle
physics, but also the possibility of having CP violation from the scalar sector was explored.
In \cite{Weinberg:1976hu} the CP properties of a model with four quarks and three Higgs bosons
were investigated. 
In the two-Higgs-doublet model (THDM) much
effort has been spent to study $\CP$ transformations; 
see for instance \cite{Gunion:2005ja, Maniatis:2007vn, Ferreira:2009wh, Ferreira:2010yh}.
The introduction of bilinears has led to an enormous simplification of
the description of any $n$-Higgs-boson-doublet model (nHDM) 
\cite{Nagel:2004sw, Maniatis:2006fs, Nishi:2006tg, Nishi:2007nh, Ivanov:2010ww, Branco:2011iw, Maniatis:2014oza, Maniatis:2015gma}.
In particular, a study of generalised CP transformations ($\CPg$) in the THDM was presented in \cite{Maniatis:2007vn} using
the method of bilinears. It turned out that these $\CPg$ transformations have a simple geometric interpretation.
In the space of the bilinears they
correspond to reflections on 
planes or to a point reflection.
A THDM has been studied in detail which is symmetric under this  $\CPg$ point reflection \cite{Maniatis:2007de,Maniatis:2009vp, Maniatis:2009by, Maniatis:2010sb}. 
This model has been shown to have 
interesting consequences: a viable model of this kind has to have at least two
fermion families with a large mass hierarchy.
In this way, a $\CPg$ symmetry gives a theoretical argument
for family replication. Recent studies of 3HDM's can, for instance, be found in~\cite{Aranda:2014jua, Chakrabarty:2015kmt, Emmanuel-Costa:2016vej, Emmanuel-Costa:2017bti, Ogreid:2017alh, Ogreid:2018xdx, Pramanick:2017wry, Cordero:2017owj, Maniatis:2015kma}.
Symmetries of the 3HDM have been studied in~\cite{Ivanov:2012fp, Keus:2013hya}.

Here we want to study generalised 
parity ($\Pg$) and charge conjugation times parity ($\CPg$) transformations for the 
three-Higgs-doublet model, 3HDM. Our paper is organised as follows. 

First we review briefly the bilinear approach for the
case of three Higgs-boson doublets. This is done in section~\ref{CPs}, whereas basis transformations
are briefly discussed in section~\ref{basis}. 
Followed by these preparations we study in section~\ref{CPg} the standard $\Ps$ and  $\CPs$ transformations and
generalised $\Pg$ and $\CPg$ transformations. In section~\ref{section5} we consider flavour
transformations of the three Higgs-boson doublets in order to bring the
generalised P and CP transformations to a standard form.
Eventually in section~\ref{section6} we classify all generalised $\Pg$ and $\CPg$ transformations by suitable 
choices of bases. 
Details of the calculation can be found in the appendices~\ref{appendixA}, \ref{appendixB}, and \ref{appendixC}. 
In appendix~\ref{appendixD} we discuss our results in the context to the conventional basis
of the potential, written in terms of scalar products of the Higgs-boson doublets. 
In appendix~\ref{appendixE} we briefly discuss the standard $\Pp$  and 
$\CP$ transformations
for the case of an arbitrary number of $n$~Higgs-boson doublets, that is, for the nHDM. 
The standard $\CPs$ transformations correspond to reflections in the space of bilinears for the THDM as well as the 3HDM.
We show that this does not hold in the nHDM for certain values of~$n$.

\section{Bilinears in the 3HDM}
\label{CPs}

We  will consider models with three Higgs-boson doublets which all 
carry the same hypercharge $y=+1/2$ and denote the
complex doublet fields by
\begin{equation}
\label{eq-doubldef}
\varphi_i(x) = 
\begin{pmatrix} 
\varphi^+_i(x) \\
 \varphi^0_i(x)
 \end{pmatrix}, \qquad i=1,2,3.
\end{equation}
We shall consider Yang-Mills-Higgs Lagrangians of the form
\begin{equation} \label{Lang}
{\cal L}(x) = {\cal L}_{\text{YM}}(x) + \sum_{i=1}^3 
\left(D_\mu \varphi_i(x) \right)^\dagger 
\left(D^\mu \varphi_i(x) \right) - V(\varphi_i),
\end{equation} 
where ${\cal L}_{\text{YM}}(x)$ is the standard Yang-Mills Lagrangian for the gauge bosons
$W_\lambda^j(x)$ ($j=$1,2,3) of $SU(2)_L$ and $B_\lambda(x)$ of $U(1)_Y$;
see for instance \cite{Nachtmann}. Furthermore, $D_\mu$ is the \eweakgroup-covariant derivative and 
$V(\varphi_i)$ is the gauge-invariant potential term. A detailed study of this type of models with 
respect to stability and symmetry breaking was presented in \cite{Maniatis:2014oza}. 
In this article we discussed in detail the bilinears for the 3HDM which will also play an essential
role in our present article. In order to make our present paper self contained we repeat here the
main points of the bilinear method for the 3HDM. 

The most general \eweakgroup gauge-invariant Higgs potential
can only be a function of products of
the Higgs-boson doublets in the form
\begin{equation}
\label{eq-potterms}
\varphi_i(x)^{\dagger}\varphi_j(x),
\qquad i,j \in \{1, 2, 3\}.
\end{equation}
We now introduce the $3 \times 2$~matrix of the Higgs-boson fields (see section 2 of \cite{Maniatis:2014oza})
\begin{equation} \label{phi}
\phi(x) = 
\begin{pmatrix} 
\varphi^+_1(x) & \varphi^0_1(x)  \\
\varphi^+_2(x) & \varphi^0_2(x)  \\
\varphi^+_3(x) & \varphi^0_3(x)
\end{pmatrix} = 
\begin{pmatrix} 
\varphi_1^\trans(x) \\
\varphi_2^\trans(x) \\
\varphi_3^\trans(x) \\
\end{pmatrix} .
\end{equation}
All possible \eweakgroup invariant
scalar products \eqref{eq-potterms} may be arranged into the hermitian $3\by 3$~matrix
\begin{gather}
\label{eq-kmat}
\twomat{K}(x) =  \phi(x) \phi^\dagger(x) =
\begin{pmatrix}
  \varphi_1^{\dagger}(x)\varphi_1(x) & \varphi_2^{\dagger}(x)\varphi_1(x) & \varphi_3^{\dagger}(x)\varphi_1(x) \\
  \varphi_1^{\dagger}(x)\varphi_2(x) & \varphi_2^{\dagger}(x)\varphi_2(x) & \varphi_3^{\dagger}(x)\varphi_2(x) \\
  \varphi_1^{\dagger}(x)\varphi_3(x) & \varphi_2^{\dagger}(x)\varphi_3(x) & \varphi_3^{\dagger}(x)\varphi_3(x)  
\end{pmatrix}.
\end{gather}

A basis for the matrices $\twomat{K}(x)$ is given by the $3 \times 3$ matrices
\begin{equation} \label{2.4a}
\lambda_\alpha, \quad \alpha = 0,1,\ldots,8 \;,
\end{equation}
where 
\begin{equation}\label{2.4b}
\lambda_0 = \sqrt{\frac{2}{3}} \unitmatrix_3
\end{equation}
is the conveniently scaled unit matrix and $\lambda_a$, $a=1,\ldots,8$
are the Gell-Mann matrices. 
Here and in
the following we will assume that greek indices ($\alpha$, $\beta$, $\ldots$) run from 0 to $8$ and 
latin indices ($a$, $b$, $\ldots$) from 1 to $8$.
We have
\begin{equation} \label{2.4c}
\tr (\lambda_\alpha \lambda_\beta) = 2 \delta_{\alpha \beta}, \qquad
\tr (\lambda_\alpha) = \sqrt{6}\; \delta_{\alpha 0}.
\end{equation}
The matrix~$\twomat{K}$~\eqref{eq-kmat} can be decomposed as 
\begin{equation} \label{2.5}
\twomat{K}(x) = \frac{1}{2} K_\alpha(x) \lambda_\alpha ,
\end{equation}
where the real coefficients $K_\alpha$, called the {\em bilinears}, are given by 
\begin{equation} \label{2.6}
K_\alpha(x) = K_\alpha^*(x) = \tr (\twomat{K}(x) \lambda_\alpha).
\end{equation}
Note that in particular
\begin{equation}
K_0(x) = \tr (\twomat{K}(x) \lambda_0)
= \sqrt{\frac{2}{3}}\left(\varphi_1^\dagger(x) \varphi_1(x) + \varphi_2^\dagger(x) \varphi_2(x) + \varphi_3^\dagger(x) \varphi_3(x) \right) .
\end{equation}
With the matrix $\twomat{K}(x)$, as defined in terms of the doublet fields
in~\eqref{eq-kmat}, as well as the decomposition~\eqref{2.5}, \eqref{2.6},
we may immediately express the scalar products in terms of the bilinears; see appendix \ref{appendixA}.
The matrix $\twomat{K}(x)$~\eqref{eq-kmat} is positive semidefinite
which follows directly from its definition $\twomat{K}(x)=\phi(x)\phi^\dagger(x)$. 
The nine coefficients  
$K_\alpha(x)$ of its decomposition~\eqref{2.5} are completely
fixed given the Higgs-boson fields.
The $3 \times 2$~matrix $\phi(x)$ has trivially
rank less than or equal to two, from which it follows that this holds also for the matrix~$\twomat{K}$.
As has been shown in detail in~\cite{Maniatis:2006fs}, (see the theorem~5 there), 
to any hermitian $3 \times 3$ matrix~$\twomat{K}(x)$ with rank less than or equal to two
there correspond Higgs-boson fields
$\varphi_i(x)$, $i=1,2,3$, which are determined uniquely, up to gauge transformations.
The bilinears parametrise the gauge orbits of the three Higgs fields \eqref{eq-doubldef}. The space of the bilinears is
the subset of the nine-dimensional space of real vectors ($K_0, \ldots, K_8$) satisfying 
\begin{equation} \label{2.100}
K_0 \ge 0,\qquad
(\tr \twomat{K})^2 - \tr \twomat{K}^2 = K_0^2 - \frac{1}{2} K_a K_a \ge 0,\qquad
\det(\twomat{K}) = \frac{1}{12} G_{\alpha \beta \gamma} K_\alpha K_\beta K_\gamma =0,
\end{equation} 
where the constants $G_{\alpha \beta \gamma}$ are given in \eqref{A.3} of appendix \ref{appendixA}
(see (2.16), (A.31), and (A.32) of \cite{Maniatis:2014oza}).  

Any 3HDM potential
leading to a renormalisable theory
 can, in terms of bilinears, be written in the form
\begin{equation} \label{Vind}
V = 
\xi_0 K_0 + \xi_a K_a + \eta_{00} K_0^2 + 
2 K_0 \eta_a K_a + K_a E_{ab} K_b
\end{equation}
with  real parameters: $\xi_0$, $\eta_{00}$, $\xi_a$, $\eta_a$ 
and the symmetric parameter matrix $E_{ab}=E_{ba}$ with $a,b=1,\ldots, 8$ \cite{Maniatis:2014oza}.
Note that a constant term in the potential can always be dropped.

Defining the eight-component vectors $\tvec{K}$, $\tvec{\xi}$, $\tvec{\eta}$,
and $8 \times 8$ matrix $E$ by
\begin{equation} \label{comp}
\tvec{K} = (K_a), \qquad
\tvec{\xi} = (\xi_a), \quad
\tvec{\eta} = (\eta_a), \quad
E = (E_{ab}),
\end{equation}
we can write the general 3HDM potential in the form
\begin{equation}
\label{VK}
V = 
\xi_0 K_0 + \tvec{\xi}^\trans \tvec{K} + \eta_{00} K_0^2 + 
2 K_0 \tvec{\eta}^\trans \tvec{K} + \tvec{K}^\trans E \tvec{K}.
\end{equation}

\section{Change of basis}
\label{basis}

Let us now study an arbitrary unitary mixing of the Higgs-boson doublets of
the form (see section 3 of \cite{Maniatis:2014oza})
\begin{align}
&\varphi_i(x) \to U_{ij} \varphi_j(x), \qquad i,j \in \{1,2,3\},
\intertext{\noindent
with $U=\left(U_{ij}\right)$ a unitary $3 \times 3$ matrix, $U \in U(3)$. 
This change of basis corresponds to the following transformations
of the $3 \times 2$ matrix $\phi(x)$ and of
the $3 \times 3$ matrix $\twomat{K}(x)$ defined in \eqref{phi} and \eqref{eq-kmat}, respectively,
}
&\phi(x) \to \phi^{(U)}(x) = U \phi(x),\\ 
&\twomat{K}(x) = \phi(x) \phi^\dagger(x) \to \twomat{K}^{(U)}(x) = U \phi(x) \phi^\dagger(x) U^\dagger = 
U \twomat{K}(x) U^\dagger .\\
\intertext{
The bilinears $K_a(x)$ transform under a change of basis as
} \label{changebasis} \nonumber
&K_0(x) \to K_0^{(U)}(x)=K_0(x),\\
&K_a(x) = \tr ( \twomat{K}(x) \lambda_a) \to K_a^{(U)}(x) = \tr ( U \twomat{K}(x) U^\dagger \lambda_a)=
R_{ab}(U)  \tr ( \twomat{K}(x) \lambda_b)= 
R_{ab}(U) K_b(x),
\end{align}
where the matrix $R(U)=\left(R_{ab}(U)\right)$ is given by
\begin{equation} \label{rotation}
U^\dagger \lambda_a U = R_{ab}(U) \lambda_b .
\end{equation}
The $8 \times 8$ matrix~\mbox{$R(U)$} has the properties
\begin{equation}
\label{eq-rprodet}
R^\ast(U)=R(U),
\qquad
R^\trans(U)\, R(U) = \unitmatrix_{8},
\qquad
\det \left( R(U) \right) = 1,
\end{equation}
that is, $R(U)\in SO(8)$.
Let us note that the $R(U)$ form only a subset of $SO(8)$.

Under the replacement~\eqref{changebasis}, the Higgs potential~\eqref{VK} remains unchanged
if we simultaneously transform the parameters as follows
\begin{equation}
\label{eq-partrafo}
\begin{alignedat}{2}
\xi_0 &\to \xi^{(U)}_0 = \xi_0,  & \tvec{\xi} &\to \tvec{\xi}^{(U)} = R(U)\,\tvec{\xi}, \\
\eta_{00} &\to \eta^{(U)}_{00}=\eta_{00}, &  \tvec{\eta} &\to \tvec{\eta}^{(U)} = R(U)\,\tvec{\eta}, \\
 E & \to E^{(U)}= R(U)\,E\,R^\trans(U).
\end{alignedat}
\end{equation}

\section{Generalised $\Pp$ and $\CP$ transformations in the 3HDM}
\label{CPg}

The standard parity transformation, $\Ps$, reads
\begin{equation} \label{4.100}
\begin{split}
\varphi_i(x) & \stackrel{\Ps}{\longrightarrow} \varphi_i(x'), \\
W_\lambda^i(x) &\stackrel{\Ps}{\longrightarrow} W^{i \lambda}(x') \\
B_\lambda(x) &\stackrel{\Ps}{\longrightarrow} B^{\lambda}(x'), 
\end{split}
\end{equation}
where $i \in \{1,2,3\}$ and 
\begin{equation} \label{4.101}
x  =
\begin{pmatrix} t\\ \tvec{x} \end{pmatrix}, \qquad
x' =
\begin{pmatrix} t\\ - \tvec{x} \end{pmatrix} .
\end{equation}
For the bilinears we find from \eqref{4.100}
\begin{equation} \label{4.102}
K_\alpha(x) \stackrel{\Ps}{\longrightarrow} K'_\alpha(x) = K_\alpha(x').
\end{equation}
The Lagrangian \eqref{Lang} is invariant under $\Ps$.
Of course, once we include fermions in the usual way, 
parity invariance is lost. But in the present article we shall consider only 
the Lagrangian \eqref{Lang} and its possible symmetries. 

Next we consider the 
standard $\CP$ transformation, $\CPs$,
\begin{equation}\label{4.103}
\begin{split}
\varphi_i(x) &\stackrel{\CPs}{\longrightarrow} \varphi'_i(x) = \varphi_i^*(x'), 
\qquad (i=1,2,3).\\
W_\lambda^1(x) &\stackrel{\CPs}{\longrightarrow} -W^{\lambda 1}(x'), \\
W_\lambda^2(x) &\stackrel{\CPs}{\longrightarrow} \phantom{+}W^{\lambda 2}(x'), \\
W_\lambda^3(x) &\stackrel{\CPs}{\longrightarrow} -W^{\lambda 3}(x'), \\
B_\lambda(x) &\stackrel{\CPs}{\longrightarrow} - B^{\lambda}(x').
\end{split}
\end{equation}
Here $x$ and $x'$ are again given by \eqref{4.101}.
For  $\phi(x)$ \eqref{phi} and $\twomat{K}(x)$ \eqref{eq-kmat} we get from \eqref{4.103}
\begin{equation} \label{4.2}
\begin{split}
\phi(x) &\stackrel{\CPs}{\longrightarrow} \phi'(x) = \phi^*(x'),\\
\twomat{K}(x)   &\stackrel{\CPs}{\longrightarrow} \twomat{K}'(x) = \twomat{K}^*(x') = \twomat{K}^\trans(x'),
\end{split}
\end{equation}
and for the bilinears \eqref{2.6}
\begin{equation} \label{4.6}
\begin{split}
K_0(x)  &\stackrel{\CPs}{\longrightarrow} K_0'(x) = K_0(x'),\\
K_a(x)  &\stackrel{\CPs}{\longrightarrow} K_a'(x) = \tr (\twomat{K}^\trans(x') \lambda_a) 
= \tr (\twomat{K}(x') \lambda_a^\trans)
= \hat{C}_{ab}^s K_b(x') \;,
\end{split}
\end{equation}
where we define the $8 \times 8$ matrix $\hat{C}^s=\left(\hat{C}^s_{ab}\right)$ by
\begin{equation} \label{RdefCPs}
\lambda_a^\trans = \hat{C}^s_{ab} \lambda_b.
\end{equation}
Explicitly we get from the Gell-Mann matrices
\begin{equation} \label{CPs3HDM}
\hat{C}^s = \left(\hat{C}^s_{ab}\right) = \diag (1, -1, 1, 1, -1, 1, -1, 1).
\end{equation}
Obviously, this matrix has  the properties
\begin{equation} \label{17.65}
\hat{C}^s = \hat{C}^{s \trans}, \qquad
\hat{C}^s \hat{C}^s = \unitmatrix_8, \qquad
\det (\hat{C}^s) = -1 \;.
\end{equation}
The $\CPs$ transformation gives, applied twice, again the 
trivial transformation in terms of the doublet fields:
\begin{equation}
\varphi_i(x)  \stackrel{\CPs \circ \CPs}{\longrightarrow} \varphi_i(x), 
\qquad i=1,2,3.
\end{equation}

In appendix~\ref{appendixC} we discuss the standard $\Pp$ and $\CP$ transformations for the
case of $n$ Higgs-boson doublets, that is, for the nHDM. 

We shall now define generalised parity ($\Pg$) and CP transformations ($\CPg$) for the 3HDM. 
We do this at the level of the bilinears with the following requirements. 
\begin{itemize}
\item[(1)] Both, $\Pg$ and $\CPg$
transformations are required to be linear in the $K_\alpha$ of the form
\begin{equation}
\begin{split} \label{17.67}
K_0(x) & \longrightarrow K'_0(x) = K_0(x'),\\
K_a(x) & \longrightarrow K'_a(x) = \hat{C}_{ab} K_b(x'),
\end{split}
\end{equation}
\begin{equation} \label{4.10}
\hat{C} = \left( \hat{C}_{ab} \right) .
\end{equation}
We require furthermore, that the length of the vector ($K_a$) is left invariant
\begin{equation} \label{4.10a}
K'_a(x) K'_a(x) = K_a(x') K_a(x') .
\end{equation}
Note that this is the case for the $\Ps$ and $\CPs$ transformations;
see \eqref{4.102}, \eqref{4.6}, and \eqref{CPs3HDM}, respectively.

\item[(2)] The allowed space of the bilinears $K_\alpha$ must
not be left. 
This requires that
the $K'_\alpha(x)$ must fulfil \eqref{2.100} if the original $K_\alpha(x)$ do so. 

\item[(3)] Application of a $\Pg$ or $\CPg$ transformation twice should
give back the original bilinears $K_\alpha(x)$. That is, we require
\begin{equation} \label{17.73}
\hat{C} \hat{C} = \unitmatrix_8.
\end{equation}
\end{itemize}

From \eqref{17.73} we see that we have 
\begin{equation} \label{4.170}
 \det (\hat{C})^2 =1, \qquad  \det (\hat{C}) =\pm 1.
\end{equation}
We shall call transformations where $\det (\hat{C}) = + 1$ generalised $\Pp$ ($\Pg$)
and where $\det (\hat{C}) = - 1$ generalised CP ($\CPg$) transformations.

We have seen in section \ref{basis} that we can make flavour $U(3)$ rotations of the Higgs fields.
If we make a corresponding transformation of the parameters of the potential we get the 
same theory but written in a different basis.

We now want to study how the matrix $\hat{C}$ of \eqref{4.10} looks like 
in a new basis.
We have under a change of basis $U$ from \eqref{changebasis} and \eqref{17.67}
\begin{equation} \label{4.200}
\begin{split}
K_0^{(U)'}(x) &= K_0^{(U)}(x'),\\
K_a^{(U)'}(x) &= R_{ab}(U) K_b'(x) = R_{ab}(U) \hat{C}_{bc} K_c(x') =
R_{ab}(U) \hat{C}_{bc} R_{cd}^{-1}(U) K_d^{(U)}(x') .
\end{split}
\end{equation}
Therefore, the matrix $\hat{C}$ of a generalised $\Pp$ or $\CP$ transformation in a new basis reads
\begin{equation} \label{4.201}
\hat{C}^{(U)} = R(U) \hat{C} R^\trans(U).
\end{equation}
Generalised transformations where the corresponding matrices $\hat{C}$ are related by a 
flavour transformation \eqref{4.201} will be called equivalent. The main purpose of 
our present article is to determine all equivalence classes of generalised parity and generalised
$\CP$ transformations, $\Pg$ and $\CPg$, respectively. 

\section{Standard forms of generalised $\Pp$ and $\CP$ transformations}
\label{section5}

The problem is now to find standard forms for the matrices $\hat{C}$ which satisfy our conditions
(1)-(3) to which general matrices $\hat{C}$ can be brought using only the flavour transformations \eqref{4.201}.

\subsection{The equations for the matrix $\hat{C}$}

We start from \eqref{17.67}, \eqref{4.10}, and \eqref{4.10a} which imply 
\begin{equation} \label{17.75}
K'_a(x) K'_a(x) = \hat{C}_{a b} K_b(x') \hat{C}_{a c} K_c(x') 
= K_b(x') \left( \hat{C}^\trans \hat{C} \right)_{bc} K_c(x')
= K_a(x') K_a(x').
\end{equation}
This is fulfilled if
\begin{equation} \label{17.76}
\hat{C}^\trans  \hat{C} = \unitmatrix_8.
\end{equation}
But since the $K_a(x')$ have to fulfil the condition \eqref{2.100}, equation \eqref{17.76}
does not follow immediately. 
We present the proof of \eqref{17.76} in appendix \ref{appendixA}. 
The technique which we use there is to consider \eqref{17.75} for
a suitable number of special cases where the model with three 
Higgs fields reduces to one with only two Higgs fields. 

Next we consider the last equation of \eqref{2.100} which must be fulfilled 
both for $K_\alpha(x)$ and $K'_\alpha(x)$ from \eqref{17.67}; see condition (2) above:
\begin{equation} \label{5.20}
G_{\alpha \beta \gamma} K_\alpha(x') K_\beta(x') K_\gamma(x') = 0,\qquad
G_{\alpha \beta \gamma} K'_\alpha(x) K'_\beta(x) K'_\gamma(x) = 0.
\end{equation}
With the explicit form of the constants $G_{\alpha \beta \gamma}$ from 
\eqref{A.3} we get from \eqref{5.20}
\begin{equation} \label{5.21}
\begin{split}
&\sqrt{\frac{2}{3}} (K_0(x'))^3 - \sqrt{\frac{3}{2}} K_0(x') K_a(x') K_a(x')
+ d_{a b c} K_a(x') K_b(x') K_c(x') = 0,\\
&\sqrt{\frac{2}{3}} (K'_0(x))^3 - \sqrt{\frac{3}{2}} K'_0(x) K'_a(x) K'_a(x)
+ d_{a b c} K'_a(x) K'_b(x) K'_c(x) = 0.
\end{split}
\end{equation}
Using now \eqref{17.67} and \eqref{4.10a} we get
\begin{equation} \label{5.22}
d_{a' b' c'} \hat{C}_{a' a} \hat{C}_{b' b} \hat{C}_{c' c}
K_a(x') K_b(x') K_c(x')
=
d_{a b c}
K_a(x') K_b(x') K_c(x') .
\end{equation}
This is satisfied if
\begin{equation} \label{5.23}
d_{a' b' c'} \hat{C}_{a' a} \hat{C}_{b' b} \hat{C}_{c' c} = d_{a b c} .
\end{equation}
Again, \eqref{5.23} does not follow immediately from \eqref{5.22}
since the $K_a(x')$ are not independent. They have to fulfil \eqref{2.100}.
The proof of \eqref{5.23} is presented in appendix \ref{appendixA}
considering \eqref{5.22} for a suitable number of special cases.

To summarise: the equations which determine the matrices
$\hat{C}$ \eqref{4.10} of a $\Pg$ or a $\CPg$ transformation are \eqref{17.73},
\eqref{17.76}, and \eqref{5.23}. 

\subsection{Flavour transformations of the matrices $\hat{C}$}

In this section we shall use the flavour transformations \eqref{4.201} to bring
the matrices $\hat{C}$ \eqref{4.10} to a standard form. 
From \eqref{17.73} and \eqref{17.76} we see that $\hat{C}$ is a symmetric matrix
\begin{equation} \label{5.24}
\hat{C}^\trans = \hat{C}.
\end{equation}
Therefore, $\hat{C}$ can be diagonalised by an $SO(8)$ matrix. Due to
\eqref{17.73} the eigenvalues of $\hat{C}$ can only be $\pm 1$. Note that,
a priori, we do not know if such an $SO(8)$ matrix diagonalising $\hat{C}$ can be 
written as a flavour transformation $R(U)$ as in \eqref{4.201}. In any case,
$\hat{C}$ has eight eigenvectors which we can, without loss of generality, assume to be real.

Suppose $\tvec{c}^{(8)}$ is one of these eigenvectors, which we 
assume to be normalised,
\begin{equation} \label{17.94}
\tvec{c}^{(8) \trans} \tvec{c}^{(8)} = c^{(8)}_a c^{(8)}_a =1.
\end{equation}
Under a basis transformation \eqref{4.201} this eigenvector
transforms as
\begin{equation} \label{17.95}
\tvec{c}^{(8)} \longrightarrow R(U) \tvec{c}^{(8)}.
\end{equation}
We use this in order to bring $\tvec{c}^{(8)}$ to a standard form.
For this we consider the matrix
\begin{equation} \label{17.97}
\Lambda^{(8)} = c^{(8)}_a \lambda_a \;.
\end{equation}
Under a basis transformation \eqref{17.95} we get
\begin{equation} \label{17.98}
\Lambda^{(8)} \longrightarrow R_{ab}(U) c^{(8)}_b \lambda_a  
= c^{(8)}_b R_{ba}^\trans(U) \lambda_a
= c^{(8)}_b R_{ba}(U^{-1}) \lambda_a
= c^{(8)}_b U^{-1 \dagger} \lambda_b U^{-1}
= U  \Lambda^{(8)} U^\dagger.
\end{equation}
We have furthermore
\begin{equation} \label{17.98a}
\Lambda^{(8)\dagger} = \Lambda^{(8)}, \qquad \tr (\Lambda^{(8)}) =0, \qquad
\tr (\Lambda^{(8)} \Lambda^{(8)} ) = 2 c^{(8)}_a c^{(8)}_a = 2.
\end{equation}
Through a basis transformation $U$ we may diagonalise $\Lambda^{(8)}$.
Taking the explicit form of the Gell-Mann matrices into account we get
\begin{equation} \label{17.100}
\left.\Lambda^{(8)}\right|_{\text{diag.}} = c_3' \lambda_3 + c_8' \lambda_8.
\end{equation}
Therefore, taking into account \eqref{17.94}, we can, by a basis change, achieve the form
\begin{equation} \label{17.101}
\tvec{c}^{(8)} = 
\begin{pmatrix}
0, & 0, & \sin(\chi), & 0, & 0, & 0, & 0, & \cos(\chi)
\end{pmatrix}^\trans .
\end{equation}
Since an overall sign of $\tvec{c}^{(8)}$ is irrelevant 
we can restrict the parameter $\chi$ to  $-\pi/2 < \chi \le \pi/2$,
corresponding to $\cos(\chi) \ge 0$. But we may further restrict
$\chi$ in the following way. 
Let us consider the matrix $\Lambda^{(8)}(\chi)$:
\begin{equation} \label{5.11}
\begin{split}
\Lambda^{(8)}(\chi) & = 
 \sin(\chi) \lambda_3 + \cos (\chi) \lambda_8
=
\frac{2}{\sqrt{3}} 
\diag \left( 
\frac{1}{2} \cos(\chi)+ \frac{\sqrt{3}}{2} \sin(\chi),
\frac{1}{2} \cos(\chi)- \frac{\sqrt{3}}{2} \sin(\chi),
- \cos(\chi)
\right)
\\ &
=
\frac{2}{\sqrt{3}} 
\diag \left( 
\cos(\chi-\frac{\pi}{3}),
\cos(\chi+\frac{\pi}{3}),
-\cos(\chi)
\right).
\end{split}
\end{equation}
Since we can, by $SU(3)$ basis transformations, exchange the 
eigenvalues, we can require that the eigenvalues of 
$\Lambda^{(8)}(\chi)$ are in decreasing order, 
that is, 
\begin{equation} \label{5.11a}
\frac{1}{2} \cos(\chi)+ \frac{\sqrt{3}}{2} \sin(\chi) \ge
\frac{1}{2} \cos(\chi)- \frac{\sqrt{3}}{2} \sin(\chi) \ge
- \cos(\chi).
\end{equation} 
From these requirements we get $0 \le \chi \le \pi/2$ and $\chi \le \pi/3$,
that is $0 \le \chi \le \pi/3$. 

We consider now the range $\pi/6 < \chi \le \pi/3$
and set 
\begin{equation}
\chi' = \frac{\pi}{3} - \chi.
\end{equation}
From \eqref{5.11} we get then with $0 \le \chi' < \pi/6$
\begin{equation} \label{5.11c}
\Lambda^{(8)}(\chi) =
-\frac{2}{\sqrt{3}} \diag
( - \cos (\chi'), \cos(\chi'+\frac{\pi}{3}), \cos(\chi'-\frac{\pi}{3}) ).
\end{equation}
Since the overall sign of $\Lambda^{(8)}$ and the order of the
eigenvalues do not matter we see that \eqref{5.11c} is
equivalent to \eqref{5.11} with $\chi$ replaced by $\chi'$. 
Taking everything together we see that by flavour 
transformations we can bring $\Lambda^{(8)}$ and
correspondingly $\tvec{c}^{(8)}$ to the forms \eqref{5.11} and \eqref{17.101}, 
respectively, with 
\begin{equation}  \label{5.12}
0 \le \chi \le \pi/6 \;.
\end{equation}

For the standard form of $\hat{C}$  we choose now
in addition to $\tvec{c}^{(8)}$ \eqref{17.101} with $\chi$ from \eqref{5.12} seven
orthonormal vectors $\tvec{c}^{(1)}$ to $\tvec{c}^{(7)}$,
which are also orthogonal to $\tvec{c}^{(8)}$:
\begin{equation}
\tvec{c}^{(i)\trans} \tvec{c}^{(j)} = \delta_{ij}, 
\qquad
\tvec{c}^{(8)\trans} \tvec{c}^{(i)} = 0,
\qquad i,j \in \{1, \ldots, 7\}.
\end{equation}
Explicitly we use 
\begin{equation}
\begin{alignedat}{2} \label{5.14}
&\tvec{c}^{(1)} = \begin{pmatrix} 1, & 0, & 0, & 0, & 0, & 0, & 0, & 0 \end{pmatrix}^\trans, 
&&\tvec{c}^{(2)} = \begin{pmatrix} 0, & 1, & 0, & 0, & 0, & 0, & 0, & 0 \end{pmatrix}^\trans, 
\\
&\tvec{c}^{(3)} = \begin{pmatrix} 0, & 0, & \cos(\chi), & 0, & 0, & 0, & 0, & -\sin(\chi) \end{pmatrix}^\trans, \quad
&&\tvec{c}^{(4)} = \begin{pmatrix} 0, & 0, & 0, & 1, & 0, & 0, & 0, & 0 \end{pmatrix}^\trans, 
\\
&\tvec{c}^{(5)}  = \begin{pmatrix} 0, & 0, & 0, & 0, & 1, & 0, & 0, & 0 \end{pmatrix}^\trans, 
&&\tvec{c}^{(6)} = \begin{pmatrix} 0, & 0, & 0, & 0, & 0, & 1, & 0, & 0 \end{pmatrix}^\trans, 
\\
&\tvec{c}^{(7)} = \begin{pmatrix} 0, & 0, & 0, & 0, & 0, & 0, & 1, & 0 \end{pmatrix}^\trans .
\end{alignedat}
\end{equation}
The matrix $\hat{C}$ has then the form
\begin{equation} \label{5.15}
\hat{C} = \sum_{i,j=1}^7 
\tvec{c}^{(i)} \tilde{C}_{ij} \tvec{c}^{(i)\trans}
+
c_8 \;
\tvec{c}^{(8)} \tvec{c}^{(8)\trans}.
\end{equation}
From \eqref{17.73} and \eqref{17.76} we must have 
\begin{equation} \label{5.16}
\tilde{C}_{ij} = \tilde{C}_{ji}, \qquad 
\tilde{C}_{ij} \tilde{C}_{jl} = \delta_{il}, \qquad
i,j \in \{1, \ldots, 7\}.
\end{equation}

We can further simplify $(\tilde{C}_{ij})$. For $\chi=0$ we have from \eqref{5.11}
\begin{equation} \label{5.17}
\Lambda^{(8)}(0) =
\frac{2}{\sqrt{3}} \diag
( \frac{1}{2}, \frac{1}{2}, -1) = \lambda_8.
\end{equation}
This matrix is invariant under the following $U(3)$ flavour transformations
\begin{equation} \label{5.18}
U = 
\begin{pmatrix}
U^{(2)} & 0\\
0 & e^{i \varphi}
\end{pmatrix},
\qquad
U^{(2)} U^{(2) \dagger} = \unitmatrix_2.
\end{equation}
That is, $U^{(2)} \in U(2)$. For the case $\chi=0$ we have 
$\tilde{C}_{ij}  = \hat{C}_{ij}$ and we can, using the flavour
transformations $R(U)$ \eqref{4.201} with $U$ from \eqref{5.18} achieve that
\begin{equation} \label{5.19}
\hat{C}_{12}=\hat{C}_{23}=\hat{C}_{13}=\hat{C}_{45}=0 , 
\qquad
\hat{C}_{33} \ge \hat{C}_{11} \ge \hat{C}_{22},
\qquad
\hat{C}_{44} \ge \hat{C}_{55};
\end{equation}
see appendix \ref{appendixB}.
For the general case, $0 < \chi \le \pi/6$, all three eigenvalues
of  $\Lambda^{(8)}(\chi)$ \eqref{5.11} are different.
Therefore, we can only make the following $U(3)$ transformations leaving
$\Lambda^{(8)}(\chi)$ invariant.
\begin{equation} \label{5.200}
U(\vartheta, \psi, \varphi) =
e^{i \vartheta}
\diag (
e^{ \frac{i}{2} \psi},
e^{ - \frac{i}{2} \psi},
e^{ i \varphi}
).
\end{equation}
With the corresponding flavour transformations $R(U)$ from \eqref{4.201} we can achieve
\begin{equation} \label{5.201}
\tilde{C}_{12} = \tilde{C}_{45} = 0, \qquad
\tilde{C}_{11} \ge \tilde{C}_{22}, \qquad 
\tilde{C}_{44 } \ge \tilde{C}_{55};
\end{equation}
see appendix \ref{appendixB}.


\section{The solutions for $\hat{C}$}
\label{section6}

In this section we give the solutions for the matrices $\hat{C}$ \eqref{4.10}.
The equations to be solved are the following: we have from \eqref{17.73}
and \eqref{17.76}
\begin{equation} \label{6.1}
\hat{C} \hat{C} = \hat{C}^\trans \hat{C} = \unitmatrix_8.
\end{equation}
Using this we can write \eqref{5.23} in the form 
\begin{equation} \label{6.2}
d_{abc} \hat{C}_{br} \hat{C}_{cs} = \hat{C}_{a a'} d_{a' rs}.
\end{equation}\\

With the help of the flavour transformations,
as explained in section \ref{section5}, we can, without 
loss of generality, assume that one eigenvector $\tvec{c}^{(8)}$ of 
$\hat{C}$ has the form \eqref{17.101} with $0 \le \chi \le \pi/6$; see \eqref{5.12}. 
We shall  first treat the case $\chi=0$, where we can transform 
$\hat{C}$ such that \eqref{5.19} holds. We have then
\begin{equation} \label{6.3}
\begin{split}
\hat{C}_{a8} &= 0, \text{ for } a = 1, \ldots, 7,\\ 
\hat{C}_{12} &= \hat{C}_{23} = \hat{C}_{13} = \hat{C}_{45} =0,\\
\hat{C}_{33} & \ge \hat{C}_{11} \ge \hat{C}_{22},\\
\hat{C}_{44} & \ge \hat{C}_{55},\\
\hat{C}_{88} &= \pm 1. 
\end{split}
\end{equation}

We shall now consider special values of $a$, $r$, $s$ in \eqref{6.2},
take into account \eqref{6.3}, and determine from this all elements $\hat{C}_{ab}$.
For $a=r=s=8$ we get from \eqref{6.2} and \eqref{6.3}
\begin{equation} \label{6.4}
d_{888} \hat{C}_{88} \hat{C}_{88} = \hat{C}_{88} d_{888}.
\end{equation}
Since $d_{888} \neq 0$, see table \ref{table1} in appendix \ref{appendixA}, we get 
\begin{equation} \label{6.5}
\hat{C}_{88} = 1.
\end{equation}
Next we set $a=8$, $r=s=1$. From \eqref{6.2}, \eqref{6.3}, and \eqref{6.5}
we get then
\begin{equation} \label{6.6}
d_{8bc} \hat{C}_{b1} \hat{C}_{c1} = d_{811},\qquad
-\frac{1}{2}
\left( \hat{C}_{41}\hat{C}_{41}+ \hat{C}_{51}\hat{C}_{51}+\hat{C}_{61}\hat{C}_{61}
+\hat{C}_{71}\hat{C}_{71} \right) =1 - \hat{C}_{11}\hat{C}_{11}.
\end{equation}
Since all eigenvalues of $\hat{C}$ are $\pm 1$ we must have
\begin{equation} \label{6.7}
-1 \le \hat{C}_{11} \le 1, \qquad \hat{C}_{11} \hat{C}_{11} \le 1.
\end{equation}
This shows that the r.h.s. and l.h.s. of \eqref{6.6} are $\ge 0$ and
$\le 0$, respectively. Therefore, both have to be zero, which implies
\begin{equation} \label{6.8}
\hat{C}_{41} = \hat{C}_{51}= \hat{C}_{61}= \hat{C}_{71}=0, \qquad
\hat{C}_{11} = \pm 1.
\end{equation}
In a similar way we show, setting in \eqref{6.2} $a=8$, $r=s=2$, and
$a=8$, $r=s=3$, that we must have 
\begin{equation} \label{6.9}
\hat{C}_{42} = \hat{C}_{52}= \hat{C}_{62}= \hat{C}_{72}=0, \qquad
\hat{C}_{22} = \pm 1,
\end{equation}
and
\begin{equation} \label{6.10}
\hat{C}_{43} = \hat{C}_{53}= \hat{C}_{63}= \hat{C}_{73}=0, \qquad
\hat{C}_{33} = \pm 1.
\end{equation}

Now we consider the cases 
\begin{equation}
\begin{split}
&a=8, r=s=4,\\
&a=8, r=s=5,\\
&a=3, r=s=4,\\
&a=3, r=s=5.
\end{split}
\end{equation} 
These give the relations
\begin{equation} \label{6.12}
\begin{split}
&(\hat{C}_{44})^2 + (\hat{C}_{64})^2 + (\hat{C}_{74})^2 = 1,\\
&(\hat{C}_{55})^2 + (\hat{C}_{65})^2 + (\hat{C}_{75})^2 = 1,\\
&(\hat{C}_{44})^2 - (\hat{C}_{64})^2 - (\hat{C}_{74})^2 = \hat{C}_{33},\\
&(\hat{C}_{55})^2 - (\hat{C}_{65})^2 - (\hat{C}_{75})^2 = \hat{C}_{33},
\end{split}
\end{equation}
with the solution
\begin{equation} \label{6.13}
\begin{split}
&(\hat{C}_{44})^2 = \frac{1}{2} (1 + \hat{C}_{33}),\\
&(\hat{C}_{55})^2 = \frac{1}{2} (1 + \hat{C}_{33}),\\
&(\hat{C}_{64})^2+(\hat{C}_{74})^2 = \frac{1}{2} (1 - \hat{C}_{33}),\\
&(\hat{C}_{65})^2+(\hat{C}_{75})^2 = \frac{1}{2} (1 - \hat{C}_{33}),
\end{split}
\end{equation}
where $\hat{C}_{33} = \pm 1$; see \eqref{6.10}.
At this point we have to distinguish two cases.

We start with the case $\hat{C}_{33} = + 1$.
From \eqref{6.13} we get then
\begin{equation} \label{6.14}
\hat{C}_{64} =\hat{C}_{74} = \hat{C}_{65} = \hat{C}_{75} =0,
\quad
\hat{C}_{44} = \pm 1, \quad 
\hat{C}_{55} = \pm 1.
\end{equation}
Choosing now in \eqref{6.2} $a=4$, $r=1$, $s=6$ and 
$a=5$, $r=1$, $s=7$ we get
\begin{equation} \label{6.15}
\hat{C}_{11} \hat{C}_{66} = \hat{C}_{44}, \quad
\hat{C}_{11} \hat{C}_{77} = \hat{C}_{55}.
\end{equation}
With \eqref{6.8}, \eqref{6.14} and the orthogonality of $\hat{C}$ this implies
\begin{equation} \label{6.16}
\hat{C}_{66} = \pm 1, 
\quad
\hat{C}_{77} = \pm 1, 
 \quad
 \hat{C}_{67} = 0. 
\end{equation}

Taking everything together we see that we have already shown that $\hat{C}$ 
must be diagonal with 
$\hat{C}_{33}= \hat{C}_{88}=1$ and
$\hat{C}_{aa}=\pm 1$ for $a=1,2,4,5,6,7$.
From \eqref{5.23} we get now that we must have
\begin{equation} \label{6.17}
\hat{C}_{aa}\hat{C}_{bb}\hat{C}_{cc} =1, \quad \text{if } d_{abc} \neq 0.
\end{equation}
The solutions of \eqref{6.17} are now easily obtained using the $d_{abc}$ values
from table \ref{table1} in appendix \ref{appendixA}. We label the solutions by (S1), \ldots, (S8); see table \ref{table2}.
There we also list the values of $\det(\hat{C})$ and of $N_\pm$, where
\begin{equation} \label{6.16a}
N_+\; (N_-) = \text{the number of eigenvalues of $\hat{C}$ equal to }
+1\; (-1)\;.
\end{equation}

We have given here the detailed derivation of the solution matrices $\hat{C}$ of \eqref{6.1}
and \eqref{6.2} for the case $\chi=0$, $\hat{C}_{33}=1$ in \eqref{6.13}. 
In appendix \ref{appendixC} we show that for $\chi=0$, $\hat{C}_{33}=-1$ in \eqref{6.13}
there is no solution. Furthermore we discuss in appendix \ref{appendixC} the cases 
with $0 < \chi \le \pi/6$. It turns out that also there only the solutions of table \ref{table2}
exist. Thus, in table \ref{table2} we have listed indeed all solutions of \eqref{6.1} and \eqref{6.2}, 
of course, apart from flavour transformations of them. 

\begin{table}
\begin{tabular}{|c|rrrrrrrr|c|c|c|}
\hline
 & $\hat{C}_{11}$ &  $\hat{C}_{22}$ & $\hat{C}_{33}$  & $\hat{C}_{44}$  
 & $\hat{C}_{55}$  & $\hat{C}_{66}$  & $\hat{C}_{77}$  & $\hat{C}_{88}$  &  $\det (\hat{C})$
 & $N_+$ & $N_-$\\
\hline
(S1) & 1 & 1 & 1& 1& 1& 1& 1& 1 &\phantom{+}\!\!1 & 8 & 0\\
(S2) & 1 & 1 & 1& -1& -1& -1& -1& 1 &\phantom{+}\!\!1 & 4 & 4\\
(S3) & 1 & -1 & 1& 1& -1& 1& -1& 1 & -1& 5 & 3\\
(S4) & 1 & -1 & 1& -1& 1& -1& 1& 1 & -1& 5 & 3\\
(S5) & -1 & 1 & 1& 1& -1& -1& 1& 1 & -1& 5 & 3\\
(S6) & -1 & 1 & 1& -1& 1& 1& -1& 1 & -1& 5 & 3\\
(S7) & -1 & -1 & 1& 1& 1& -1& -1& 1 &\phantom{+}\!\!1& 4 & 4\\
(S8) & -1 & -1 & 1& -1& -1& 1& 1& 1 &\phantom{+}\!\!1& 4 & 4\\
\hline
\end{tabular}
\caption{\label{table2} All solutions $\hat{C}$ of \eqref{6.1} and \eqref{6.2} can be 
brought, with suitable flavour transformations, to diagonal form. Listed are the diagonal
elements $\hat{C}_{aa}$ ($a=1, \ldots, 8$) and $\det(\hat{C})$ for the solutions
(S1), \ldots, (S8) and the numbers $N_+$ and $N_-$ of eigenvalues $+1$ and $-1$, 
respectively.}
\end{table}
 
We shall now discuss the meaning of the solutions (S1), \ldots, (S8) from table \ref{table2}.
That is, we will discuss the transformations \eqref{17.67}
\begin{equation} \label{6.18}
\begin{split}
K_0(x) &\longrightarrow K'_0(x) = K_0(x'),\\
K_a(x) &\longrightarrow K'_a(x) = \hat{C}_{ab} K_b(x'),
\quad
\hat{C} = \diag ( \hat{C}_{11}, \ldots, \hat{C}_{88} ),
\end{split}
\end{equation}
for the $\hat{C}_{aa}$ from table \ref{table2}.
In the following the solution matrix for (Si) will be labeled $\hat{C}^{(i)}$, 
$i = 1, \dots, 8$.

\begin{itemize}
\item[(S1)] Here $\hat{C}^{(1)} = \unitmatrix_8$.
Inserting this in \eqref{6.18} we see that we get the
standard parity transformation $\Ps$; see \eqref{4.100}-\eqref{4.102}.


\item[(S2)] 
Here $\det(\hat{C}^{(2)}) = +1$. Inserting $\hat{C}^{(2)}$ in \eqref{6.18} we obtain 
a generalised parity transformation $\Pg$ which, at the field level, reads
\begin{equation}
\begin{pmatrix} \varphi_1(x) \\ \varphi_2(x) \\ \varphi_3(x) \end{pmatrix}
\longrightarrow
\begin{pmatrix} \phantom{+}\varphi_1(x') \\ \phantom{+}\varphi_2(x') \\ -\varphi_3(x') \end{pmatrix}.
\end{equation}
This is the standard parity transformation followed by a flavour transformation,
see \eqref{B.4},
\begin{equation} \label{6.20}
U(0,0,\pi) = \diag ( 1, 1, -1).
\end{equation}
Since the sets of eigenvalues $\hat{C}_{aa}$ for (S1) and (S2) are different,
$\Ps$ and the $\Pg$ above are inequivalent.

\item[(S3)] 
This case corresponds to the standard CP transformation $\CPs$; see 
\eqref{4.2}-\eqref{CPs3HDM}. That is, $\hat{C}^{(3)} \equiv \hat{C}^s$.

\item[(S4)] 
This case corresponds to a standard CP transformation followed by a flavour transformation \eqref{6.20}
\begin{equation} \label{6.21}
\begin{pmatrix} \varphi_1(x) \\ \varphi_2(x) \\ \varphi_3(x) \end{pmatrix}
\longrightarrow
\begin{pmatrix} \phantom{+}\varphi_1^*(x') \\ \phantom{+}\varphi_2^*(x') \\ -\varphi_3^*(x') \end{pmatrix}.
\end{equation}

But $\hat{C}^{(4)}$ is equivalent to the standard CP transformation $\hat{C}^{(3)}$.
We have with \eqref{B.4}, \eqref{B.5}, and \eqref{4.201}
\begin{equation} \label{6.22}
\hat{C}^{(4)} =
R( U (0, 0, \pi/2)) \hat{C}^{(3)} R^\trans( U (0, 0, \pi/2)).
\end{equation}

\item[(S5)]
This corresponds to the standard CP transformation followed by a flavour transformation
$U (-\pi/2, \pi, \pi/2)$  from \eqref{B.4}. We get
\begin{equation} \label{6.23}
\begin{pmatrix} \varphi_1(x) \\ \varphi_2(x) \\ \varphi_3(x) \end{pmatrix}
\longrightarrow
\begin{pmatrix} \phantom{+}\varphi_1^*(x') \\ -\varphi_2^*(x') \\ \phantom{+}\varphi_3^*(x') \end{pmatrix}.
\end{equation}
Also this generalised CP transformation is equivalent to the standard one since we have
\begin{equation} \label{6.24}
\hat{C}^{(5)} =
R( U (0, \pi/2, \pi/4)) \hat{C}^{(3)} R^\trans( U (0, \pi/2, \pi/4) ).
\end{equation}

\item[(S6)]
Here we have a standard CP transformation followed by a flavour transformation
$U (-\pi/2, \pi, -\pi/2)$; see \eqref{B.4}. We get
\begin{equation} \label{6.25}
\begin{pmatrix} \varphi_1(x) \\ \varphi_2(x) \\ \varphi_3(x) \end{pmatrix}
\longrightarrow
\begin{pmatrix} \phantom{+}\varphi_1^*(x') \\ -\varphi_2^*(x') \\ -\varphi_3^*(x') \end{pmatrix}.
\end{equation}
Also $\hat{C}^{(6)}$ is equivalent to the standard CP transformation since we have,
see \eqref{B.5},
\begin{equation} \label{6.26}
\hat{C}^{(6)} =
R( U (0, \pi/2, -\pi/4)) \hat{C}^{(3)} R^\trans( U (0, \pi/2, -\pi/4) ).
\end{equation}

\item[(S7)] 
This case corresponds to the standard parity transformation followed by a flavour transformation
$U (-\pi/2, \pi, \pi/2)$; see \eqref{B.4}. We get
\begin{equation} \label{6.27}
\begin{pmatrix} \varphi_1(x) \\ \varphi_2(x) \\ \varphi_3(x) \end{pmatrix}
\longrightarrow
\begin{pmatrix} \phantom{+}\varphi_1(x') \\ -\varphi_2(x') \\ \phantom{+}\varphi_3(x') \end{pmatrix}.
\end{equation}
This is equivalent to the generalised parity transformation $\Pg$ from (S2) 
since we have
\begin{equation} \label{6.28}
\hat{C}^{(7)} =
R( U ) \hat{C}^{(2)} R^\trans( U )
\end{equation}
with 
\begin{equation} \label{6.29}
U = 
\begin{pmatrix} 1 & \phantom{+}0 & 0 \\ 0 & \phantom{+}0 & 1\\ 0 & -1 & 0 \end{pmatrix}.
\end{equation}

\item[(S8)] 
Here we have a standard parity transformation followed by a flavour transformation
$U (\pi/2, \pi, -\pi/2)$ from \eqref{B.4}
\begin{equation} \label{6.272}
\begin{pmatrix} \varphi_1(x) \\ \varphi_2(x) \\ \varphi_3(x) \end{pmatrix}
\longrightarrow
\begin{pmatrix} -\varphi_1(x') \\ \phantom{+}\varphi_2(x') \\ \phantom{+}\varphi_3(x') \end{pmatrix}.
\end{equation}
Also here we find equivalence to the generalised parity transformation from (S2) 
since we have
\begin{equation} \label{6.31}
\hat{C}^{(8)} =
R( U ) \hat{C}^{(2)} R^\trans( U )
\end{equation}
with 
\begin{equation} \label{6.32}
U = 
\begin{pmatrix} \phantom{+}0 & \phantom{+}0 & \phantom{+}1 \\
\phantom{+} 0 & \phantom{+}1 & \phantom{+}0\\
 -1 & \phantom{+}0 & \phantom{+}0 \end{pmatrix}.
\end{equation}
\end{itemize}

To summarise: we have found that for the 3HDM there are two 
equivalence classes of generalised parity transformations. 
Convenient representatives of these classes are the standard 
parity transformation with the matrix $\hat{C}^{(1)}$ from (S1) and
the generalised parity transformation with the matrix $\hat{C}^{(8)}$
from (S8); see table \ref{table2}. 
All generalised CP transformations form only one equivalence class with
the standard CP transformation as representative; see
$\hat{C}^{(3)}$ from (S3) in table \ref{table2}.
In this way we have obtained a complete answer to the question of generalised P and
CP transformations in the 3HDM.

\section{Invariant potentials}
\label{invariant}
We consider now the potential of the 3HDM in the form 
\eqref{Vind} respectively \eqref{VK}. Note that all
parameters \eqref{comp} of the potential, 
written in this form, must be real.
We consider now a generalised parity ($\Pg$) or CP transformation ($\CPg$)
satisfying the conditions (1), (2), (3) of section \ref{CPg}; see
\eqref{17.67}-\eqref{17.73}. The potential $V$ is invariant under this transformation if
\begin{equation} \label{7.1}
\left. V(x) \right|_{K_\alpha'(x)} = \left. V(x') \right|_{K_\alpha(x')}.
\end{equation}
Writing this out we get
\begin{multline} \label{7.2}
K_0'(x) \xi_0 + \tvec{K}'^\trans(x) \tvec{\xi} +
(K_0'(x))^2 \eta_{00} + 2 K_0'(x) \tvec{K}'^\trans(x) \tvec{\eta} 
+ \tvec{K}'^\trans(x) E \tvec{K}'(x) =\\
K_0(x') \xi_0 + \tvec{K}^\trans(x') \tvec{\xi} +
(K_0(x'))^2 \eta_{00} + 2 K_0(x') \tvec{K}^\trans(x') \tvec{\eta} 
+ \tvec{K}^\trans(x') E \tvec{K}(x'), \qquad
\end{multline}
\begin{multline} \label{7.3}
\tvec{K}^\trans(x') \hat{C}^\trans \tvec{\xi} +
2 K_0(x') \tvec{K}^\trans(x') \hat{C}^\trans \tvec{\eta}
+ \tvec{K}^\trans(x') \hat{C}^\trans E \hat{C} \tvec{K}(x') =\\
\tvec{K}^\trans(x') \tvec{\xi} +
2 K_0(x') \tvec{K}^\trans(x') \tvec{\eta}
+ \tvec{K}^\trans(x') E \tvec{K}(x'). \qquad
\end{multline}
This must hold for all allowed $K_\alpha(x')$.
From this it follows, using \eqref{A.24} and \eqref{A.25},
that the potential $V$ of 
\eqref{Vind}, \eqref{VK} is invariant under the $\Pg$ or $\CPg$ 
transformation considered if and only if
\begin{equation} \label{7.4}
\tvec{\xi} = \hat{C} \tvec{\xi}, \quad
\tvec{\eta} = \hat{C} \tvec{\eta}, \quad
E = \hat{C} E \hat{C}^\trans \;.
\end{equation}

Let us now investigate what this implies for the cases (S1), (S3), and (S8)
of table \ref{table2}.

For the standard P transformation (S1) we have $\hat{C}^{(1)}=\unitmatrix_8$
and the conditions \eqref{7.4} are trivially fulfilled. As already mentioned in section \ref{CPg}, any potential is invariant under the standard P transformation.

If the potential $V$ allows a generalised CP invariance then there is a basis where the CP transformation has the standard form, that is, (S3) in table \ref{table2}.
In this basis we obtain from \eqref{7.4} the conditions
\begin{equation} \label{7.5}
\begin{split}
&\xi_a = 0 \qquad \text{for } a \in \{2,5,7\},\\
&\eta_a = 0 \qquad \text{for } a \in \{2,5,7\},\\
& E_{ab}=E_{ba}=0 \qquad \text{for } a \in \{2,5,7\},\; b \in \{1,3,4,6,8\}.
\end{split}
\end{equation}
In appendix \ref{appendixD} we discuss the relation of these conditions to statements on CP violation using the conventional form of the basis.

Finally we consider the generalised P transformations $\Pg$ of the class 
with representative (S8). A potential $V$ allows such a generalised $\Pg$
invariance if and only if there is a basis where we have, inserting $\hat{C}^{(8)}$ 
in \eqref{7.4} 
\begin{equation} \label{7.6}
\begin{split}
&\xi_a = 0 \qquad \text{for } a \in \{1,2,4,5\},\\
&\eta_a = 0 \qquad \text{for } a \in \{1,2,4,5\},\\
& E_{ab}=E_{ba}=0 \qquad \text{for } a \in \{1,2,4,5\}, \; b \in \{3,6,7,8\}.
\end{split}
\end{equation}

In this way we have obtained a complete overview of the invariance conditions
of the potential for generalised P and CP transformations.

\section{Conclusions} \label{conclusions}
In this paper we have considered the three-Higgs-doublet model (3HDM)
with Higgs and gauge fields only.
We have investigated generalised P and CP transformations in this model.
We have shown that there are two equivalence classes (with respect to flavour
transformations) of generalised P transformations and only one class of CP transformations. Convenient representatives for these classes are given in 
table~\ref{table2}: the standard P transformation (S1), the generalised P transformation (S8), and the standard CP transformation (S3). We have discussed the conditions which a potential has to fulfil in order to be invariant under any of these transformations. In all our work we made use of the method of bilinears.
The relation to the conventional basis for the potential is discussed in detail in appendix~\ref{appendixD}.

To summarise, we have investigated the general 3HDM in view of generalised P and CP transformations which applied twice give back the unit transformation up to possible gauge transformations. Our work gives a complete overview of such transformations and of their consequences for the potential.

A very interesting type of ``CP'' transformations was considered in~\cite{Ivanov:2015mwl}. There one has to apply the transformation four times
in order to get back the unit transformation. Of course, also such transformations can be analysed with the methods developed in the present paper. This will be dealt with in a separate work.

Finally we have made some remarks on the standard P and CP transformations in the nHDM in appendix~\ref{appendixE}.
We have shown that in the space of bilinears the standard CP transformation $\CPs$ is again given by a linear transformation of 
the corresponding bilinears with an $(n^2-1) \times (n^2-1)$ matrix $\hat{C}$ which is given explicitly. 
$\hat{C}$ is diagonal with diagonal elements $\pm 1$. The number of $-1$ elements is $n(n-1)/2$.

\appendix
\section{Proof of equations \eqref{17.76} and \eqref{5.23}}
\label{appendixA}
Here we recall some formulae for the bilinears of the 3HDM and the THDM. Then we prove equations \eqref{17.76} and \eqref{5.23}.

The bilinears of the 3HDM as defined in \eqref{2.6} are given explicitly by
\begin{alignat}{2} \nonumber
K_0 & = \sqrt{\frac{2}{3}}\left(\varphi_1^\dagger \varphi_1 + \varphi_2^\dagger \varphi_2 + \varphi_3^\dagger \varphi_3 \right),\quad
&& K_1  =
\varphi_1^\dagger \varphi_2 + \varphi_2^\dagger \varphi_1,\\ \nonumber
K_2 & = \frac{1}{i} \left(
\varphi_1^\dagger \varphi_2 - \varphi_2^\dagger \varphi_1 \right),
&&K_3  =
\varphi_1^\dagger \varphi_1 - \varphi_2^\dagger \varphi_2,\\ \label{A.1}
K_4 & =
\varphi_1^\dagger \varphi_3 + \varphi_3^\dagger \varphi_1,
&&K_5  = \frac{1}{i} \left(
\varphi_1^\dagger \varphi_3 - \varphi_3^\dagger \varphi_1 \right),\\  \nonumber
K_6 & =
\varphi_2^\dagger \varphi_3 + \varphi_3^\dagger \varphi_2,
&&K_7  = \frac{1}{i} \left(
\varphi_2^\dagger \varphi_3 - \varphi_3^\dagger \varphi_2 \right),\\ \nonumber
K_8 & = \frac{1}{\sqrt{3}}
\left(\varphi_1^\dagger \varphi_1 + \varphi_2^\dagger \varphi_2 - 2 \varphi_3^\dagger \varphi_3 \right)
\end{alignat}
The $K_\alpha$ satisfy \eqref{2.100} where $G_{\alpha \beta \gamma}$ is given
by (see (A.31) of \cite{Maniatis:2014oza})

 \begin{equation} \label{A.2}
 G_{\alpha \beta \gamma}  = \frac{1}{4} 
 \bigg\{
 \tr(\lambda_\alpha) \tr(\lambda_\beta) \tr(\lambda_\gamma)
 + \tr(\lambda_\alpha \lambda_\beta \lambda_\gamma + \lambda_\alpha \lambda_\gamma \lambda_\beta)
 - \tr(\lambda_\alpha) \tr(\lambda_\beta \lambda_\gamma)
 - \tr(\lambda_\beta) \tr(\lambda_\gamma \lambda_\alpha)
 - \tr(\lambda_\gamma) \tr(\lambda_\alpha \lambda_\beta)
 \bigg\}.
 \end{equation}
 which is completely symmetric in
 $\alpha$, $\beta$, $\gamma$. Explicitly 
we get
\begin{equation} \label{A.3}
G_{0 \beta \gamma} = \sqrt{\frac{3}{2}} \delta_{\beta 0} \delta_{\gamma 0}
- \frac{1}{\sqrt{6}} \delta_{\beta \gamma},
\qquad
G_{a b c} = d_{a b c}.
\end{equation}
The $d_{abc}$ are the usual symmetric constants of $SU(3)$.
We list the non-zero elements of the $d_{abc}$ in table \ref{table1}; see for instance \cite{Nachtmann}.


\begin{table}
\begin{tabular}{*{17}{p{25pt}}}
\hline
$a$ & 1 & 1 & 1 & 2 & 2 & 2 & 3 & 3 & 3 & 3 & 3 &4 & 5 & 6 & 7 &8\\
$b$ & 1 & 4 & 5 & 2 &4 & 5& 3 & 4 & 5 & 6 & 7 & 4 & 5 & 6 & 7 & 8\\
$c$ & 8 & 6 & 7 & 8 & 7 & 6 & 8 & 4 & 5 & 6 & 7 & 8 & 8 & 8 & 8 & 8\\
\hline
 $d_{abc}$ & $\frac{1}{\sqrt{3}}$ & $\frac{1}{2}$ & $\frac{1}{2}$ & $\frac{1}{\sqrt{3}}$ & $\!\!\!\!\!-\frac{1}{2}$  & $\frac{1}{2}$ & $\frac{1}{\sqrt{3}}$ & $\frac{1}{2}$ &  $\frac{1}{2}$ &  $\!\!\!\!\!-\frac{1}{2}$ &  $\!\!\!\!\!-\frac{1}{2}$ & $\!\!\!\!\!-\frac{1}{2\sqrt{3}}$ & $\!\!\!\!\!-\frac{1}{2\sqrt{3}}$ & $\!\!\!\!\!-\frac{1}{2\sqrt{3}}$ & $\!\!\!\!\!-\frac{1}{2\sqrt{3}}$ & $\!\!\!\!\!-\frac{1}{\sqrt{3}}$\\
\hline
\end{tabular}
\caption{\label{table1} 
The nonzero elements of the SU(3) constants $d_{abc}$.
}
\end{table}

In the following we also need the bilinears for the THDM given in \cite{Nagel:2004sw, Maniatis:2006fs}.
Therefore, we reproduce some results of these references here. 
Let 
\begin{equation} \label{A.4}
\psi_i(x) =
\begin{pmatrix} \psi_i^+ (x) \\ \psi_i^0(x) \end{pmatrix}, \quad i=1,2,
\end{equation}
be two Higgs-doublet fields with hypercharge $y=1/2$. We define
the matrix $\twomat{L}$ by 
\begin{equation} \label{A.5}
\twomat{L} =
\begin{pmatrix} 
\psi^\dagger_1 \psi_1 & \psi^\dagger_2 \psi_1\\
\psi^\dagger_1 \psi_2 & \psi^\dagger_2 \psi_2
\end{pmatrix}
\end{equation}
and the bilinears of the THDM by 
\begin{equation} \label{A.6}
\twomat{L}= \frac{1}{2}
\left( L_0 \unitmatrix_2 +
L_1 \sigma_1 + L_2 \sigma_2 + L_3 \sigma_3 \right)
\end{equation}
where $\sigma_i$ are the Pauli matrices. Explicitly we get
\begin{equation}  \label{A.7}
 \begin{split}
 L_0 &= \psi_1^\dagger \psi_1 + \psi_2^\dagger \psi_2,\\
 L_1 &= \psi_1^\dagger \psi_2 + \psi_2^\dagger \psi_1,\\
 L_2 &= \frac{1}{i} \left( \psi_1^\dagger \psi_2 - \psi_2^\dagger \psi_1 \right),\\
 L_3 &= \psi_1^\dagger \psi_1 - \psi_2^\dagger \psi_2,
 \end{split}
 \end{equation}
and from this
\begin{equation}  \label{A.8}
\begin{split}
\psi_1^\dagger \psi_1 &= \frac{1}{2} (L_0 + L_3), 
\qquad 
\psi_1^\dagger \psi_2 = \frac{1}{2} (L_1 + i L_2),\\
\psi_2^\dagger \psi_2 &= \frac{1}{2} (L_0 - L_3), 
\qquad 
\psi_2^\dagger \psi_1 = \frac{1}{2} (L_1 - i L_2);
\end{split}
\end{equation}
see section 3 of \cite{Maniatis:2006fs}. To distinguish in our present article 
3HDM and THDM quantities  we use $\psi$ and $L$ for the fields
and bilinears of the THDM, respectively, instead of $\varphi$ and
$K$ in \cite{Maniatis:2006fs}. The bilinears $L_\alpha$ satisfy
\begin{equation}  \label{A.9}
L_0 \ge 0, \qquad 
L_0^2 - L_1^2 - L_2^2 - L_3^2 \ge 0;
\end{equation}
see (36) of \cite{Maniatis:2006fs}. 

Now we are in the position to prove \eqref{17.76}. 
We have for all $K_\alpha$ satisfying \eqref{2.100} from \eqref{17.75}
\begin{equation} \label{A.10}
K_b G_{bc} K_c = K_a K_a,
\end{equation}
where
\begin{equation} \label{A.11}
G_{bc} = \left( \hat{C}^\trans \hat{C} \right)_{bc} = G_{cb}.
\end{equation}
We want to show that $G_{bc}=\delta_{bc}$. The technique for
this is to use special cases, corresponding to THDM fields in \eqref{A.10}.

First we set

\begin{alignat}{2} \label{A.12}
&(a) \qquad & \varphi_1 = \psi_1, \quad
\varphi_2 = \psi_2, \quad
\varphi_3 = 0.
\end{alignat}
This gives, using \eqref{A.7}, \eqref{A.8}, and \eqref{A.1},
\begin{equation} \label{A.13}
K_1 = L_1,\quad K_2= L_2, \quad K_3 = L_3,\qquad
K_4 = K_5 = K_6= K_7 = 0,\qquad
K_8 = \frac{1}{\sqrt{3}} L_0,
\end{equation}
and from \eqref{A.10}
\begin{equation} \label{A.14}
\begin{split}
&G_{11} L_1^2 + G_{22} L_2^2 + G_{33} L_3^2 + \frac{1}{3} G_{88} L_0^2 \\
&+ 2 G_{12} L_1 L_2 + 2 G_{13} L_1 L_3 + \frac{2}{\sqrt{3}} G_{18} L_1 L_0\\
&+ 2 G_{23} L_2 L_3 + \frac{2}{\sqrt{3}} G_{28} L_2 L_0 + \frac{2}{\sqrt{3}} G_{38} L_3 L_0 
\\
&= L_1^2 + L_2^2 + L_3^2 + \frac{1}{3} L_0^2. 
\end{split}
\end{equation}
Since the $L_0$, $L_i$ only have to satisfy \eqref{A.9} the polynomials on the left- and right-hand sides of \eqref{A.14} must be equal. This implies
\begin{equation} \label{A.15}
G_{11}=G_{22}=G_{33}=G_{88}=1,\qquad
G_{12}=G_{13}=G_{18}=G_{23}=G_{28}=G_{38}=0.
\end{equation}
We consider then six more special cases
\begin{alignat}{4} \label{A.16}
&(b) \qquad && \varphi_1 = \psi_1, \quad
&&\varphi_2 =  0, \quad
&&\varphi_3 = \psi_2, 
\\ \label{A.17}
&(c) \qquad && \varphi_1 = 0, \quad
&&\varphi_2 =  \psi_1, \quad
&&\varphi_3 = \psi_2,
\\ \label{A.18}
&(d) \qquad && \varphi_1 = \psi_1, \quad
&&\varphi_2 =  \psi_2, \quad
&&\varphi_3 = \pm \psi_2,
\\ \label{A.19}
&(e) \qquad && \varphi_1 = \psi_1, \quad
&&\varphi_2 =  \psi_2, \quad
&&\varphi_3 = \pm i \psi_2.
\end{alignat}
Proceeding in each case as shown explicitly for case $(a)$
we obtain that we must have 
\begin{equation} \label{A.20}
G_{ab} = \delta_{ab}
\end{equation}
which was to be proven. Here, all special cases $(a)$-$(e)$ can easily be treated by hand, as we have shown by \eqref{A.14}, \eqref{A.15}. But we also have written a computer program to deal with these cases. 

To prove \eqref{5.23} we proceed in the same way. We write \eqref{5.22} as
\begin{equation} \label{A.21}
D_{abc} K_a K_b K_c = d_{abc} K_a K_b K_c,
\end{equation}
where
\begin{equation} \label{A.212}
D_{abc} = d_{a'b'c'} \hat{C}_{a'a} \hat{C}_{b'b} \hat{C}_{c'c}
\end{equation}
Clearly, $D_{abc}$ is completely symmetric. We insert now again special cases, corresponding to THDMs, in \eqref{A.21} and compare the polynomials in $L_0$,
$L_1$, $L_2$, $L_3$ which we obtain on the right- and left-hand sides of \eqref{A.21}. 
The special cases which we use here are again (a) to (e) from \eqref{A.12}, \eqref{A.16}-\eqref{A.19}.
In addition we need here the choices
\begin{alignat}{4} \label{choices}
&(f) \qquad && \varphi_1 = \psi_2, \quad
&&\varphi_2 =  -i \psi_2, \quad
&&\varphi_3 = \psi_1,\\
&(g) \qquad && \varphi_1 = \psi_2, \quad
&&\varphi_2 =  \psi_1, \quad
&&\varphi_3 = 0,\\
&(h) \qquad && \varphi_1 = \psi_2, \quad
&&\varphi_2 =  0, \quad
&&\varphi_3 = \psi_1.
\end{alignat}
The comparison of the two sides of \eqref{A.21} for all these special cases is then
done with the help of a computer program and gives
\begin{equation} \label{A.23}
D_{abc} = d_{abc}.
\end{equation}
This proves \eqref{5.23}.

In a similar way we can prove the following.
Suppose that we are given real quantities $b_\alpha$ and $B_{\alpha \beta}=B_{\beta \alpha}$.
If for all allowed $K_\alpha$ we have 
\begin{equation} \label{A.24}
b_\alpha K_\alpha =0, \qquad
K_\alpha B_{\alpha \beta} K_\beta =0,
\end{equation}
then we must have 
\begin{equation} \label{A.25}
b_\alpha =0, \qquad B_{\alpha \beta} =0.
\end{equation}
For the proof we use the special cases (a)-(f) as well as the choices
\begin{alignat}{4} \label{A.26}
&(i) \qquad && \varphi_1 = \psi_1, \quad
&&\varphi_2 =  \psi_2, \quad
&&\varphi_3 = \psi_1 \pm \psi_2, 
\\ \label{A.27}
&(j) \qquad && \varphi_1 = \psi_1, \quad
&&\varphi_2 =  \psi_2, \quad
&&\varphi_3 = i \psi_1+ i \psi_2\;.
\end{alignat}

\section{Special flavour transformations}
\label{appendixB}

The flavour transformation \eqref{5.18} with $\varphi=0$, $U^{(2)} \in U(2)$,
\begin{equation} \label{B.1}
U = \begin{pmatrix} U^{(2)} & 0 \\ 0 & 1 \end{pmatrix}
\end{equation}
gives from \eqref{rotation}
\begin{equation} \label{B.2}
\left(R_{ab}(U)\right) = 
\text{
\begin{tabular}{|m{14pt}||m{45pt} m{45pt} m{45pt} | m{45pt} m{45pt} m{45pt} m{45pt} |m{12pt}|}
\hline
a\textbackslash b &\quad 1 &\quad 2 &\quad 3 &\qquad 4 &\qquad 5 &\qquad 6 &\qquad 7 & 8\\
 \hline\hline
 1 & & & & & & & &\\
 2 & \multicolumn{3}{c|}{$R^{(2)}_{ab}(U^{(2)})$}  & \multicolumn{4}{c|}{0} & 0\\ 
3 & & & & & & & &\\
 \hline
 4 & & & &  $\phantom{+}\re( U^{(2)}_{11})$ & $\phantom{+}\im( U^{(2)}_{11})$ & $\phantom{+}\re( U^{(2)}_{12})$ & 
 $\phantom{+}\im( U^{(2)}_{12})$ & \\
 5 & \multicolumn{3}{c|}{0} & $-\im( U^{(2)}_{11})$ & $\phantom{+}\re( U^{(2)}_{11})$ & $-\im( U^{(2)}_{12})$ & 
 $\phantom{+}\re( U^{(2)}_{12})$ & 0 \\
6 & & & & $\phantom{+}\re( U^{(2)}_{21})$ & $\phantom{+}\im( U^{(2)}_{21})$ & $\phantom{+}\re( U^{(2)}_{22})$ & $\phantom{+}\im( U^{(2)}_{22})$ & \\
7 & & & & $-\im( U^{(2)}_{21})$ & $\phantom{+}\re( U^{(2)}_{21})$ & $-\im( U^{(2)}_{22})$ & 
$\phantom{+}\re( U^{(2)}_{22})$ & \\
\hline
8 & \multicolumn{3}{c|}{0}  & \multicolumn{4}{c|}{0} & 1\\
\hline
\end{tabular}
}
\end{equation}
Here $R^{(2)}_{ab}(U^{(2)})$ is defined by 
\begin{equation} \label{B.3}
U^{(2)\dagger} \sigma_a U^{(2)} = R_{ab}^{(2)} \left ( U^{(2)} \right) \sigma_b,
\qquad a,b \in \{1,2,3\},
\end{equation}
where $\sigma_a$ are the Pauli matrices. The matrix 
$R^{(2)} \left ( U^{(2)} \right) = \left ( R_{ab}^{(2)} \left ( U^{(2)} \right) \right)$
is an $SO(3)$ transformation.

Next we consider a diagonal matrix from $U(3)$, see \eqref{5.200},
\begin{equation} \label{B.4}
U(\vartheta, \psi, \varphi) = e^{i \vartheta}
\begin{pmatrix}
e^{\frac{i}{2} \psi} & 0 & 0 \\
0 & e^{-\frac{i}{2} \psi} & 0 \\
0 & 0 & e^{i \varphi}
\end{pmatrix}.
\end{equation}
We get then from \eqref{rotation}
\begin{equation} \label{B.5}
\left(R_{ab} \left ( U \left(\vartheta, \psi, \varphi  \right) \right) \right) \!\!= \!\!\!
\text{
\begin{tabular}{|c||p{45pt} p{45pt} p{45pt} | p{55pt} p{55pt} | p{55pt} p{55pt} |r|}
\hline
a \textbackslash b & \quad 1 &\quad 2 &\quad 3 &\qquad 4 &\qquad 5 &\qquad 6 &\qquad 7 & 8\\
 \hline\hline
 1 & $\phantom{+}\cos(\psi)$ & $\phantom{+}\sin(\psi)$ & \quad 0 &\qquad 0 & \qquad0 &\qquad 0 &\qquad 0 & 0\\
 2 & $-\sin(\psi)$ & $\phantom{+}\cos(\psi)$  & \quad 0 & \qquad 0 & \qquad 0 & \qquad 0 & \qquad 0 & 0\\
 3 & \quad 0 & \quad 0 & \quad 1 & \qquad 0 & \qquad 0 & \qquad 0 & \qquad 0 & 0\\
 \hline
 4 & \quad 0 & \quad 0 & \quad 0 &  $\phantom{+}\cos(\frac{\psi}{2}-\varphi)$ & $\phantom{+}\sin(\frac{\psi}{2}-\varphi)$ & \qquad 0 & \qquad 0 & 0\\ 
 5 & \quad 0 & \quad 0 & \quad 0 &  $-\sin(\frac{\psi}{2}-\varphi)$ & $\phantom{+}\cos(\frac{\psi}{2}-\varphi)$ & \qquad 0 & \qquad 0 & 0\\ 
 6 & \quad 0 & \quad 0 & \quad 0 & \qquad  0 & \qquad 0 & $\phantom{+}\cos(\frac{\psi}{2}+\varphi)$ & $-\sin(\frac{\psi}{2}+\varphi)$ & 0\\ 
 7 & \quad 0 & \quad 0 & \quad 0 & \qquad  0 & \qquad 0 & $\phantom{+}\sin(\frac{\psi}{2}+\varphi)$ & $\phantom{+}\cos(\frac{\psi}{2}+\varphi)$ & 0\\ 
 \hline
8 & \quad 0 & \quad 0 & \quad 0  & \qquad 0 & \qquad 0 & \qquad 0 & \qquad 0 & 1\\
\hline
\end{tabular}
}
\end{equation}

Consider now the case $\chi=0$ from \eqref{5.12}, \eqref{5.17}. With a flavour transformation \eqref{5.18}
with $\varphi=0$ we can diagonalise the $3 \times 3$ submatrix $(\hat{C}_{ab})$, ($(1 \le a, b \le 3$) of $\hat{C}$ using
\eqref{B.2} in \eqref{4.201}.
We can also achieve the ordering of the diagonal elements $\hat{C}_{11}$, $\hat{C}_{22}$, $\hat{C}_{33}$ as 
given in \eqref{5.19}. Then we can use a diagonal matrix $U(0, 0, \varphi)$ from \eqref{B.4} to achieve $\hat{C}_{45}=0$
and $\hat{C}_{44} \ge \hat{C}_{55}$ using \eqref{B.5} in \eqref{4.201}. 

For the case $0 < \chi \le \pi/6$ we can only use $U(\vartheta, \psi, \varphi)$ \eqref{5.200}, \eqref{B.4}, for a further simplification of $\hat{C}$,
respectively $\tilde{C}$ \eqref{5.15}. Note that $R \left ( U \left(\vartheta, \psi, \varphi  \right) \right)$ \eqref{B.5}
leaves the $(3,8)$ subspace invariant. Therefore, we can apply the flavour transformation \eqref{B.5} also directly 
to $\tilde{C}$ and get from this, without loss of generality, the restrictions \eqref{5.201}. 


\section{Details of the calculation for the solutions $\hat{C}$}
\label{appendixC}
Here we complete the discussion of the solutions of \eqref{6.1} and \eqref{6.2} for the
matrices $\hat{C}$.
\begin{itemize}
\item[{\bf C1}] {\bf The case $\chi=0$, $\hat{C}_{33} = -1$}

Here we consider the case $\chi=0$ in \eqref{5.12}. We have treated this case generally up to \eqref{6.13}
where we had to distinguish two cases, $\hat{C}_{33} = \pm 1$. The case $\hat{C}_{33}= +1$ is discussed in 
section \ref{section6}. Here we discuss the case $\hat{C}_{33}= -1$. From \eqref{6.3} and \eqref{6.12} we
get then, using here and in the following the $d_{abc}$ values from table \ref{table1},
\begin{equation}  \label{C.1}
\hat{C}_{11} = \hat{C}_{22} = \hat{C}_{33} = -1, \qquad \hat{C}_{44} = \hat{C}_{55} =0.
\end{equation}
Now we set in \eqref{6.2} $a=4$, $r=1$, $s=6$ and $a=5$, $r=1$, $s=7$. This gives, 
taking into account \eqref{6.3} to \eqref{6.10}
\begin{equation} \label{C.2}
\hat{C}_{11} \hat{C}_{66} =0, \qquad \hat{C}_{11} \hat{C}_{77} =0,
\end{equation}
and, therefore, with \eqref{C.1}
\begin{equation} \label{C.3}
\hat{C}_{66} = \hat{C}_{77} =0.
\end{equation}
Next we choose $a=2$, $r=5$, $s=6$ in \eqref{6.2}. This gives with \eqref{C.1}
\begin{equation} \label{C.4}
\begin{split}
&d_{2bc} \hat{C}_{b5} \hat{C}_{c6} = \hat{C}_{22} d_{256} = - d_{256},\\
&d_{247} \hat{C}_{45} \hat{C}_{76} + d_{247} \hat{C}_{75} \hat{C}_{46}
+ d_{256} \hat{C}_{55} \hat{C}_{66}
+ d_{265} \hat{C}_{65} \hat{C}_{56} = -d_{256}.
\end{split}
\end{equation}
Using \eqref{5.19} and \eqref{C.1} this gives
\begin{equation} \label{C.5}
\hat{C}_{75} \hat{C}_{46} = 1 + (\hat{C}_{56})^2.
\end{equation}
But since $\hat{C}$ is an orthogonal matrix we have 
\begin{equation} \label{C.6}
\left| \hat{C}_{75} \hat{C}_{46} \right| \le 1
\end{equation}
and we get from \eqref{C.5}
\begin{equation} \label{C.7}
\hat{C}_{56} = 0, \qquad  \hat{C}_{75} \hat{C}_{46} =1 .
\end{equation}
Using again the orthogonality property of $\hat{C}$ we find
\begin{equation} \label{C.8}
\hat{C}_{75} = \hat{C}_{46} = \pm 1, \qquad
\hat{C}_{47} = \hat{C}_{67} = 0.
\end{equation}
Finally we choose $a=1$, $r=4$, $s=6$ in \eqref{6.2} and find
\begin{equation} \label{C.9}
d_{1bc} \hat{C}_{b4} \hat{C}_{c6} = \hat{C}_{11} d_{146} = - d_{146}, \qquad
\left( \hat{C}_{46} \right)^2 = -1.
\end{equation}
This is a contradiction and shows that there is no solution for $\hat{C}$ for
the case $\chi=0$, $\hat{C}_{33} = -1$.

\item[{\bf C2}] {\bf The case $0 < \chi < \pi/6$}

Let us next discuss the case $0 < \chi < \pi/6$ in \eqref{5.12}.
Here we make an $SO(8)$ basis transformation of $\hat{C}$.
We set 
\begin{equation} \label{C.10}
S =
\begin{pmatrix}
1 & 0 & 0 & 0 & 0 & 0 & 0 & 0\\
0 & 1 & 0 & 0 & 0 & 0 & 0 & 0\\
0 & 0 & c & 0 & 0 & 0 & 0 & -s\\
0 & 0 & 0 & 1 & 0 & 0 & 0 & 0\\
0 & 0 & 0 & 0 & 1 & 0 & 0 & 0\\
0 & 0 & 0 & 0 & 0 & 1 & 0 & 0\\
0 & 0 & 0 & 0 & 0 & 0 & 1 & 0\\
0 & 0 & s & 0 & 0 & 0 & 0 & c
\end{pmatrix}.
\end{equation}
Here and in the following we set
\begin{equation} \label{C.10a}
s = \sin(\chi), \qquad c = \cos(\chi).
\end{equation}
We emphasise that $S$ will in general {\em not} be a 
flavour transformation $R(U)$ as in \eqref{rotation}, \eqref{eq-rprodet}.
From \eqref{C.10} we have
\begin{equation} \label{C.11}
S S^\trans = \unitmatrix_8.
\end{equation}
Now we transform $\hat{C}$ with $S$ and set 
\begin{equation} \label{C.12}
\hat{H} = S \hat{C} S^\trans .
\end{equation}
We have from \eqref{6.1} and \eqref{C.11}
\begin{equation} \label{C.13}
\hat{H}^\trans = \hat{H}, \qquad
\hat{H} \hat{H} = \hat{H}^\trans \hat{H} = \unitmatrix_8.
\end{equation}
Form \eqref{6.2} we get the following condition for $\hat{H}$
\begin{equation} \label{C.14}
\tilde{d}_{a b' c'} \hat{H}_{b' b} \hat{H}_{c' c} = \hat{H}_{a a'} \tilde{d}_{a'bc},
\end{equation}
where
\begin{equation} \label{C.15}
\tilde{d}_{abc} = S_{a a'} S_{b b'} S_{c c'} d_{a' b' c'}.
\end{equation}
In table \ref{table3} we list the non-zero elements of $\tilde{d}_{abc}$.
\begin{table}
\begin{tabular}{c|cc||c|cc}
\hline
$a\; b\; c$ & \multicolumn{2}{c||}{$\tilde{d}_{abc}$} & $a\; b\; c$ & \multicolumn{2}{c}{$\tilde{d}_{abc}$} \\
\hline
  & general $\chi$ & $\chi = \pi/6$ & &  general $\chi$ & $\chi = \pi/6$ \\
\hline
3\; 3\; 3 & $-\frac{1}{\sqrt{3}} s (4 c^2-1)$ & $-\frac{1}{\sqrt{3}}$ &  6\; 6\; 3 & $-\frac{1}{2\sqrt{3}}  (\sqrt{3} c-s)$ & $-\frac{1}{2\sqrt{3}}$\\
8\; 8\; 8 & $-\frac{1}{\sqrt{3}} c (4 c^2-3)$ & $0$ &  6\; 6\; 8 & $-\frac{1}{2\sqrt{3}}  (c+\sqrt{3} s)$ & $-\frac{1}{2}$\\
1\; 1\; 3 & $-\frac{1}{\sqrt{3}} s$ & $-\frac{1}{2\sqrt{3}}$ &  7\; 7\; 3 & $-\frac{1}{2\sqrt{3}}  (\sqrt{3} c-s)$ & $-\frac{1}{2\sqrt{3}}$\\
1\; 1\; 8 & $\frac{1}{\sqrt{3}} c$ & $\frac{1}{2}$ &  7\; 7\; 8 & $-\frac{1}{2\sqrt{3}}  (c+\sqrt{3} s)$ & $-\frac{1}{2}$\\
2\; 2\; 3 & $-\frac{1}{\sqrt{3}} s$ & $-\frac{1}{2\sqrt{3}}$ &  8\; 8\; 3 & $\frac{1}{\sqrt{3}} s (4 c^2-1)$ & $\frac{1}{\sqrt{3}}$\\
2\; 2\; 8 & $\frac{1}{\sqrt{3}} c$ & $\frac{1}{2}$ &  1\; 4\; 6 & $\frac{1}{2}$ & $\frac{1}{2}$\\
3\; 3\; 8 & $\frac{1}{\sqrt{3}} c(2c^2-1)$ & $\frac{1}{4}$ &  1\; 5\; 7 & $\frac{1}{2}$ & $\frac{1}{2}$\\
4\; 4\; 3 & $\frac{1}{2\sqrt{3}} (\sqrt{3}c+s)$ & $\frac{1}{\sqrt{3}}$ &  2\; 4\; 7 & $-\frac{1}{2}$ & $-\frac{1}{2}$\\
4\; 4\; 8 & $-\frac{1}{2\sqrt{3}} (c-\sqrt{3}s)$ & $0$ &  2\; 5\; 6 & $\frac{1}{2}$ & $\frac{1}{2}$\\
5\; 5\; 3 & $\frac{1}{2\sqrt{3}} (\sqrt{3}c+s)$ & $\frac{1}{\sqrt{3}}$\\
5\; 5\; 8 & $-\frac{1}{2\sqrt{3}} (c-\sqrt{3}s)$ & $0$\\
\end{tabular}
\caption{\label{table3} The non-zero elements of $\tilde{d}_{abc}$ of \eqref{C.15} as function of $\chi$ and for the
special value $\chi=\pi/6$. Here $s=\sin(\chi)$, $c=\cos(\chi)$. 
}
\end{table}

By construction of $S$ \eqref{C.10} and from \eqref{5.201} we have
\begin{alignat}{3}
&\hat{H}_{a8} = 0 \text{ for } a=1,\ldots,7 \;,  \nonumber
\qquad && \hat{H}_{12} = \hat{H}_{45} = 0, \label{C.17}\\
& \hat{H}_{11} \ge \hat{H}_{22}, && \hat{H}_{44} \ge \hat{H}_{55},\\
&\hat{H}_{88} = \pm 1. \nonumber
\end{alignat}

Now we choose special values for $a$, $b$, $c$ in \eqref{C.14} in order to determine 
the possible solutions of this equation. We start with $a=b=c=8$. This gives with \eqref{C.17}
\begin{equation} \label{C.18}
\tilde{d}_{888} \hat{H}_{88} \hat{H}_{88} = \hat{H}_{88} \tilde{d}_{888} \;.
\end{equation}
For $0 < \chi < \pi/6$ we have $\tilde{d}_{888} \neq 0$; see table \ref{table3}. Therefore we find
\begin{equation} \label{C.19}
\hat{H}_{88} = 1.
\end{equation}
Next we choose in \eqref{C.14} $a=8$, $b=c=1$. This gives
\begin{equation} \label{C.20}
\tilde{d}_{8 b' c'} \hat{H}_{b' 1} \hat{H}_{c' 1} = \hat{H}_{88} \tilde{d}_{811} = \tilde{d}_{811}  \;,
\end{equation}
\begin{equation} \label{C.21}
-\frac{1}{2\sqrt{3}} (c- \sqrt{3}s) \big[ (\hat{H}_{41})^2 + (\hat{H}_{51})^2 \big]
-\frac{1}{2\sqrt{3}} (c+ \sqrt{3}s) \big[ (\hat{H}_{61})^2 + (\hat{H}_{71})^2 \big]
=
\frac{1}{\sqrt{3}} c \big[1 - (\hat{H}_{11})^2 - (2c^2-1) (\hat{H}_{31})^2 \big] \;.
\end{equation}
For $0 < \chi < \pi/6$ we have
\begin{equation} \label{C.22}
c-\sqrt{3} s >0, \qquad 
c+\sqrt{3} s >0, \qquad 
1 > 2 c^2 -1 > \frac{1}{2}.
\end{equation}
Furthermore we have, since $\hat{H}$ is an orthogonal matrix (see \eqref{C.13}),
\begin{equation} \label{C.23}
0 \le (\hat{H}_{11})^2 + (2c^2-1) (\hat{H}_{31})^2 \le (\hat{H}_{11})^2 + (\hat{H}_{31})^2 \le 1 \;.
\end{equation}
Therefore, the r.h.s. of \eqref{C.21} is greater than or equal to zero, the l.h.s. less than or equal to zero.
Thus, both sides must be zero and we get
\begin{equation} \label{C.24}
\hat{H}_{41} = \hat{H}_{51} = \hat{H}_{61} = \hat{H}_{71} =0,
\end{equation}
\begin{equation} \label{C.25}
(\hat{H}_{11})^2 + (2c^2-1) (\hat{H}_{31})^2 =1.
\end{equation}
Taking into account \eqref{C.24} and $\hat{H}_{81}=\hat{H}_{21}=0$ from \eqref{C.17} we get
from the orthogonality relation for $\hat{H}$
\begin{equation} \label{C.26}
(\hat{H}_{11})^2 + (\hat{H}_{31})^2 =1.
\end{equation}
From \eqref{C.25} and \eqref{C.26} we get with \eqref{C.22}
\begin{equation} \label{C.28}
\hat{H}_{31} = 0, \qquad \hat{H}_{11} = \pm 1 \;.
\end{equation}
Next we choose  $a=8$, $b=c=2$ in \eqref{C.14}.
Here the argumentation is as for $a=8$, $b=c=1$ above and we find
\begin{equation} \label{C.29}
\hat{H}_{32} = \hat{H}_{42} = \hat{H}_{52} = \hat{H}_{62}= \hat{H}_{72} =0, \qquad \hat{H}_{22} = \pm 1 \;.
\end{equation}
From the case  $a=8$, $b=c=3$ in \eqref{C.14} we get
\begin{equation} \label{C.30}
-\frac{1}{2\sqrt{3}} (c- \sqrt{3}s) \big[ (\hat{H}_{43})^2 + (\hat{H}_{53})^2 \big]
-\frac{1}{2\sqrt{3}} (c+ \sqrt{3}s) \big[ (\hat{H}_{63})^2 + (\hat{H}_{73})^2 \big]
=
\frac{c}{\sqrt{3}}  (2c^2-1)\big[1 - (\hat{H}_{33})^2 \big] \;.
\end{equation}
With \eqref{C.22} we conclude from \eqref{C.30}
\begin{equation} \label{C.31}
\hat{H}_{43} = \hat{H}_{53} = \hat{H}_{63} = \hat{H}_{73} =0, \qquad \hat{H}_{33} = \pm 1 \;.
\end{equation}

Next we consider  $a=8$, $b=c=4$ and 
$a=3$, $b=c=4$ in \eqref{C.14}. We get then
\begin{equation} \label{C.32}
\begin{split} 
 &(c- \sqrt{3}s) (\hat{H}_{44})^2 + (c+\sqrt{3} s)  \big[ (\hat{H}_{64})^2 + (\hat{H}_{74})^2 \big]
 = c - \sqrt{3} s,\\
&
 (\sqrt{3}c+s) (\hat{H}_{44})^2 - (\sqrt{3} c-s)  \big[ (\hat{H}_{64})^2 + (\hat{H}_{74})^2 \big]
=
 (\sqrt{3}c+s) \hat{H}_{33}.
 \end{split}
 \end{equation}
From \eqref{C.32} we find
\begin{equation} \label{C.33}
(\hat{H}_{44})^2 = \frac{1}{2} (1+ \hat{H}_{33}) - \frac{2sc}{\sqrt{3}}(1- \hat{H}_{33}) \;.
\end{equation}
According to \eqref{C.31} we have $\hat{H}_{33}=\pm 1$. But $\hat{H}_{33}=-1$ leads to a 
contradiction in \eqref{C.33}. Thus we must have
\begin{equation} \label{C.34}
\hat{H}_{33} = +1, \qquad (\hat{H}_{44})^2 = 1.
\end{equation}
Looking back at \eqref{C.32} this implies
\begin{equation} \label{C.35}
\hat{H}_{64} =  \hat{H}_{74} =0.
\end{equation}

Now we choose  $a=8$, $b=c=5$ and 
$a=3$, $b=c=5$ in \eqref{C.14}. This gives
\begin{equation} \label{C.36}
\begin{split}
& (c- \sqrt{3}s) (\hat{H}_{55})^2 + (c+\sqrt{3} s)  \big[ (\hat{H}_{65})^2 + (\hat{H}_{75})^2 \big]
 = c -\sqrt{3} s,  \\
 &(\sqrt{3}c+s) (\hat{H}_{55})^2 - (\sqrt{3} c-s)  \big[ (\hat{H}_{65})^2 + (\hat{H}_{75})^2 \big]
=
 (\sqrt{3}c+s) \hat{H}_{33}.
 \end{split}
\end{equation}
As above we conclude from \eqref{C.36} 
\begin{equation}  \label{C.37}
(\hat{H}_{55})^2 = 1, \qquad  \hat{H}_{65}  = \hat{H}_{75} =0\;. 
\end{equation}

From $a=4$, $b=1$, $c=6$ in \eqref{C.14} we get 
\begin{equation} \label{C.37a}
\hat{H}_{11} \hat{H}_{66} = \hat{H}_{44},
\end{equation}
\begin{equation}  \label{C.38}
 \hat{H}_{66}  =  \hat{H}_{11}   \hat{H}_{44}  = \pm 1, 
 \end{equation}
\begin{equation} \label{C.39}
 \hat{H}_{76}  =  0.
 \end{equation}
 
 Finally, from $a=5$, $b=1$, $c=7$ in \eqref{C.14} we get 
\begin{equation} \label{C.40}
 \hat{H}_{77}  =  \hat{H}_{11}   \hat{H}_{55}  = \pm 1.
 \end{equation}

Collecting now everything together we have shown that $\hat{H}$ has diagonal form with
\begin{equation} \label{C.41}
 \hat{H}_{33}  =    \hat{H}_{88}  =  1.
 \end{equation}
But now we can transform back to $\hat{C}$ using \eqref{C.12} and we find
\begin{equation} \label{C.42}
\hat{C} = S^\trans \hat{H} S = \diag ( \hat{C}_{11}, \hat{C}_{22}, \hat{C}_{33}, \hat{C}_{44}, \hat{C}_{55}, \hat{C}_{66}, \hat{C}_{77}, \hat{C}_{88})
\end{equation}
with
\begin{equation} \label{C.43}
\hat{C}_{33} = \hat{C}_{88} = 1, \qquad
C_{ii} = \pm 1, \text{ for } i = 1,2,4,5,6,7.
\end{equation}
From there on we can follow the analysis as in section \ref{section6} from \eqref{6.16} onwards.
We find then also here exactly the same solutions (S1) to (S8) listed in table \ref{table2}.

\item[{\bf C3}] {\bf The case $\chi = \pi/6$}

Here we start as in section C2, \eqref{C.10} to \eqref{C.18}. But here $\hat{d}_{888} =0$, see table \ref{table3}, 
so we can only conclude here
\begin{equation} \label{C.44}
\hat{H}_{88} = \pm 1 .
\end{equation}

Let us discuss first the case $\hat{H}_{88} = +1$. We set in \eqref{C.14} 
 $a=8$, $b=c=1$. This gives
\begin{equation} \label{C.45}
 -\big[(\hat{H}_{61})^2 + (\hat{H}_{71})^2 \big]   = 1 - \big[(\hat{H}_{11})^2 + \frac{1}{2}(\hat{H}_{31})^2 \big].
 \end{equation}
 Since $\hat{H}$ is an orthogonal matrix, see \eqref{C.12}, \eqref{C.13}, the r.h.s of \eqref{C.45} is $\ge 0$. Therefore,
 both sides of \eqref{C.45} must be zero which implies
\begin{equation} \label{C.46}
 \hat{H}_{61}  = \hat{H}_{71} = 0,
 \end{equation}
\begin{equation} \label{C.47}
(\hat{H}_{11})^2 + \frac{1}{2} (\hat{H}_{31})^2 =1.
\end{equation}
Using again the orthogonality property of $\hat{H}$, \eqref{C.17} and \eqref{C.46}, we get
\begin{equation} \label{C.48}
(\hat{H}_{11})^2 +  (\hat{H}_{31})^2  +  (\hat{H}_{41})^2  +  (\hat{H}_{51})^2 =1.
\end{equation}
From \eqref{C.47} and \eqref{C.48} we get 
\begin{equation} \label{C.49}
\frac{1}{2}(\hat{H}_{31})^2 +  (\hat{H}_{41})^2  +  (\hat{H}_{51})^2 =0, \qquad
 \hat{H}_{31}  = \hat{H}_{41}= \hat{H}_{51} = 0, \qquad
 \hat{H}_{11} = \pm 1 .
\end{equation}
Next we consider $a=8$, $b=c=2$ in \eqref{C.14}. This gives
\begin{equation} \label{C.50}
 -\big[(\hat{H}_{62})^2 + (\hat{H}_{72})^2 \big]   = 1 - \big[(\hat{H}_{22})^2 + \frac{1}{2}(\hat{H}_{32})^2 \big].
 \end{equation}
As above we conclude here
\begin{equation} \label{C.51}
 \hat{H}_{32}  = \hat{H}_{42}= \hat{H}_{52}= \hat{H}_{62}= \hat{H}_{72} = 0, \qquad
 \hat{H}_{22} = \pm 1 .
\end{equation}
The choice $a=8$, $b=c=3$ in \eqref{C.14} leads to
\begin{equation} \label{C.52}
 -2 \big[(\hat{H}_{63})^2 + (\hat{H}_{73})^2 \big]   = 1 -  (\hat{H}_{33})^2
 \end{equation}
which implies
\begin{equation} \label{C.53}
 \hat{H}_{43}  = \hat{H}_{53}= \hat{H}_{63}= \hat{H}_{73} = 0, \qquad
 \hat{H}_{33} = \pm 1 .
\end{equation}
From $a=8$, $b=c=4$ in \eqref{C.14} we get
\begin{equation} \label{C.54}
 (\hat{H}_{64})^2 + (\hat{H}_{74})^2 = 0, \qquad
 \hat{H}_{64} = \hat{H}_{74} = 0.
 \end{equation}
Together with \eqref{C.17}, \eqref{C.49}, \eqref{C.51}, and \eqref{C.53}, this implies
\begin{equation} \label{C.55}
 \hat{H}_{44} = \pm 1 .
\end{equation}
From $a=8$, $b=c=5$ in \eqref{C.14} we get
\begin{equation} \label{C.56}
 (\hat{H}_{65})^2 + (\hat{H}_{75})^2 = 0, \qquad
 \hat{H}_{65} = \hat{H}_{75} = 0.
 \end{equation}
Together with \eqref{C.17}, \eqref{C.49}, \eqref{C.51}, and \eqref{C.53}, this implies
\begin{equation} \label{C.57}
 \hat{H}_{55} = \pm 1 .
\end{equation}
From $a=3$, $b=c=4$ in \eqref{C.14} we get
\begin{equation} \label{C.58}
 (\hat{H}_{44})^2 = \hat{H}_{33}
\end{equation}
and, therefore, with \eqref{C.55}
\begin{equation} \label{C.59}
 \hat{H}_{33} = 1 .
\end{equation}
From $a=4$, $b=1$, $c=6$ in \eqref{C.14} we get
\begin{equation} \label{C.60}
 \hat{H}_{11} \hat{H}_{66} = \hat{H}_{44}
\end{equation}
which implies with \eqref{C.49} and \eqref{C.55}
\begin{equation} \label{C.61}
 \hat{H}_{66} = \pm 1.
\end{equation}
All relations found so far plus the orthogonality property of $\hat{H}$ imply now also 
\begin{equation} \label{C.62}
 \hat{H}_{76}= 0, \qquad \hat{H}_{77} = \pm 1.
\end{equation}
Collecting everything together we have found here that $\hat{H}$ is a diagonal matrix with 
 $\hat{H}_{33}=\hat{H}_{88}= 1$, exactly as found 
 in \eqref{C.41}. Using the reasoning as in \eqref{C.42}, \eqref{C.43} we find also here again exactly the solutions
 (S1) to (S8) from table~\ref{table2}.\\

Finally we have to discuss the case $\chi=\pi/6$, 
 $\hat{H}_{88} = -1$. Here we set 
 $a=8$, $b=c=1$ in \eqref{C.14} and find
 \begin{equation} \label{C.63}
 (\hat{H}_{11})^2 + \frac{1}{2} (\hat{H}_{31})^2 =
 -\big[ 1 -  (\hat{H}_{61})^2 -  (\hat{H}_{71})^2 \big].
 \end{equation}
 This implies
 \begin{equation} \label{C.64}
  \hat{H}_{11} = \hat{H}_{31} = 0, \qquad
 (\hat{H}_{61})^2 + (\hat{H}_{71})^2 = 1, \qquad
  \hat{H}_{41} = \hat{H}_{51} = 0.
 \end{equation}
We can, therefore, set
\begin{equation} \label{C.65}
\hat{H}_{61} = \cos (\alpha), \qquad
\hat{H}_{71} = \sin (\alpha), \quad 0 \le \alpha < 2 \pi.
\end{equation}
Next we set $a=8$, $b=c=2$ in \eqref{C.14} and get
 \begin{equation} \label{C.66}
 (\hat{H}_{22})^2 + \frac{1}{2} (\hat{H}_{32})^2 =
 -\big[ 1 -  (\hat{H}_{62})^2 -  (\hat{H}_{72})^2 \big].
 \end{equation}
 which implies
 \begin{equation} \label{C.67}
  \hat{H}_{22} = \hat{H}_{32} = 0, \qquad
 (\hat{H}_{62})^2 + (\hat{H}_{72})^2 = 1, \qquad
  \hat{H}_{42} = \hat{H}_{52} = 0.
 \end{equation}
 We set
\begin{equation} \label{C.68}
\hat{H}_{62} = \cos (\beta), \qquad
\hat{H}_{72} = \sin (\beta), \quad 0 \le \beta < 2 \pi.
\end{equation}
The orthogonality of $\hat{H}$ implies now
\begin{equation} \label{C.69}
\hat{H}_{61}\hat{H}_{62} +\hat{H}_{71}\hat{H}_{72} =0, \qquad
\cos(\alpha) \cos(\beta) + \sin(\alpha) \sin(\beta) = \cos(\alpha -\beta) = 0,
\quad
\alpha -\beta = \pm \pi/2.
\end{equation}
The orthogonality relations also imply
\begin{equation} \label{C.70}
\hat{H}_{61}\hat{H}_{63} +\hat{H}_{71}\hat{H}_{73} =0, \qquad
\hat{H}_{62}\hat{H}_{63} +\hat{H}_{72}\hat{H}_{73} =0.
\end{equation}
Since
\begin{equation} \label{C.71}
\hat{H}_{61}\hat{H}_{72} -\hat{H}_{71}\hat{H}_{62}=
\cos(\alpha) \sin(\beta) -  \sin(\alpha) \cos(\beta) = \sin(\beta - \alpha) =
\sin(\mp \pi/2) \neq 0
\end{equation}
we can conclude from \eqref{C.70}
\begin{equation} \label{C.72}
 \hat{H}_{63} = \hat{H}_{73} = 0.
\end{equation}
In a completely analogous way we find
\begin{equation} \label{C.73}
 \hat{H}_{64} = \hat{H}_{74} = 0, \qquad
 \hat{H}_{65} = \hat{H}_{75} = 0, \qquad
 \hat{H}_{66} = \hat{H}_{76} = 0, \qquad
 \hat{H}_{67} = \hat{H}_{77} = 0.
\end{equation}
Finally we  choose $a=8$, $b=c=3$ in \eqref{C.14} which gives
\begin{equation} \label{C.74}
\tilde{d}_{8 b' c'} \hat{H}_{b' 3} \hat{H}_{c' 3} = \hat{H}_{88} \tilde{d}_{833} = - \tilde{d}_{833}  \;.
\end{equation}
With \eqref{C.17}, \eqref{C.64}, \eqref{C.67}, \eqref{C.72}, and table \ref{table3} we get from \eqref{C.74}
\begin{equation} \label{C.75}
 (\hat{H}_{33})^2 = -1.
\end{equation}
This is a contradiction and shows that there is no solution of \eqref{C.14} for $\chi=\pi/6$ and $\hat{H}_{88} = -1$. 
\end{itemize}

\section{The potential in the conventional basis}
\label{appendixD}
In our paper we have always worked with the potential expressed as a polynomial in the bilinears $K_\alpha$ \eqref{2.6};
see \eqref{Vind} and \eqref{VK}. In this case all the parameters \eqref{comp} of the potential are necessarily real.
But frequently the potential is written as a polynomial in the field products $\varphi_1^\dagger \varphi_1$,
$\varphi_2^\dagger \varphi_2$, etc.; see for instance \cite{Gunion:2005ja}. Then the parameters of 
this polynomial for the potential need not all be real. In the following we shall discuss the connection of these
two ways of writing the potential. We shall also discuss how the conditions of CP invariance look like in such a basis.

We start by writing the transformation \eqref{A.1} from the products of the Higgs fields to the $K_\alpha$ in matrix form.
For this we introduce the 9 dimensional vector
\begin{equation} \label{D.1}
\tilde{P}^\trans = 
\begin{pmatrix}
\varphi_1^\dagger \varphi_1, &
\varphi_2^\dagger \varphi_2, &
\varphi_3^\dagger \varphi_3, &
\varphi_1^\dagger \varphi_2, &
\varphi_2^\dagger \varphi_1, &
\varphi_1^\dagger \varphi_3, &
\varphi_3^\dagger \varphi_1, &
\varphi_2^\dagger \varphi_3, &
\varphi_3^\dagger \varphi_2
\end{pmatrix}.
\end{equation}
We have then from \eqref{A.1}
\begin{equation} \label{D.2}
\tilde{K} = \begin{pmatrix} K_0 \\ \tvec{K} \end{pmatrix} = \tilde{A} \tilde{P}
\end{equation}
with the $9 \times 9$ matrix $\tilde{A}$ given by
\begin{equation} \label{D.3}
\tilde{A} =
\begin{pmatrix} 
\sqrt{\frac{2}{3}} & \sqrt{\frac{2}{3}} &\sqrt{\frac{2}{3}} & 0 &0 &0 &0 &0 &0 \\
0 &0 &0 & 1 &1 & 0 &0 &0 &0 \\
0 &0 &0 & -i &i & 0 &0 &0 &0 \\
1 &-1 &0 & 0 &0 & 0 &0 &0 &0 \\
0 &0 &0 & 0 &0 & 1 &1 &0 &0 \\
0 &0 &0 & 0 &0 & -i &i &0 &0 \\
0 &0 &0 & 0 &0 & 0 &0 &1 &1\\
0 &0 &0 & 0 &0 & 0 &0 &-i &i\\
\frac{1}{\sqrt{3}} & \frac{1}{\sqrt{3}} &-\frac{2}{\sqrt{3}} & 0 &0 &0 &0 &0 &0
\end{pmatrix} .
\end{equation}
The reverse transformation reads
\begin{equation} \label{A.103}
\tilde{P}  = \tilde{A}^{-1}   \begin{pmatrix} K_0 \\ \tvec{K} \end{pmatrix} = \tilde{A}^{-1} \tilde{K}
\end{equation}
with 
\begin{equation} \label{D.4}
\tilde{A}^{-1} =
\begin{pmatrix} 
\frac{1}{\sqrt{6}} &0 & 0 & \frac{1}{2} & 0 &0 &0 &0 &   \frac{1}{2\sqrt{3}} \\
\frac{1}{\sqrt{6}} &0 & 0 & -\frac{1}{2} & 0 &0 &0 &0 &   \frac{1}{2\sqrt{3}} \\
\frac{1}{\sqrt{6}} &0 & 0 & 0 & 0 &0 &0 &0 &   -\frac{1}{\sqrt{3}} \\
0 &\frac{1}{2} &\frac{i}{2} & 0 &0 & 0 &0 &0 &0 \\
0 &\frac{1}{2} &-\frac{i}{2} & 0 &0 & 0 &0 &0 &0 \\
0 & 0 &0 & 0 &\frac{1}{2} &\frac{i}{2}  &0 &0 &0 \\
0 & 0 &0 & 0 &\frac{1}{2} &-\frac{i}{2}  &0 &0 &0 \\
0 & 0 &0 & 0 &0 & 0 &\frac{1}{2} &\frac{i}{2} &0 \\
0 & 0 &0 & 0 &0 & 0 &\frac{1}{2} &-\frac{i}{2} &0
\end{pmatrix} .
\end{equation}
We introduce now from \eqref{Vind} to \eqref{VK}
\begin{equation} \label{D.5}
\tilde{\xi} =  \begin{pmatrix} \xi_0 \\ \tvec{\xi} \end{pmatrix}, \qquad
\tilde{E} =
 \begin{pmatrix} \eta_{00} & \tvec{\eta}^\trans \\ \tvec{\eta}  & E  \end{pmatrix} = \tilde{E}^\trans.
\end{equation}
With this we can write the potential $V$ as follows
\begin{equation} \label{D.6}
V= \frac{1}{2} \tilde{K}^\trans \tilde{\xi} +
 \frac{1}{2}  \tilde{\xi}^\trans \tilde{K} + 
 \tilde{K}^\trans \tilde{E} \tilde{K} =
 \frac{1}{2} \tilde{P}^\dagger \tilde{\zeta} +
 \frac{1}{2}  \tilde{\zeta}^\dagger \tilde{P} + 
 \tilde{P}^\dagger \tilde{F} \tilde{P}
 \end{equation}
 where
 \begin{equation} \label{D.7}
 \tilde{\zeta} = \tilde{A}^\dagger \tilde{\xi}, \qquad
 \tilde{F} = \tilde{A}^\dagger \tilde{E} \tilde{A}.
 \end{equation}
Note that $\tilde{\zeta}$ and $\tilde{F}$ will have imaginary parts. Indeed,
we split $\tilde{A}$ \eqref{D.3} into its real ($\tilde{A}_R$) and imaginary ($\tilde{A}_I$) parts,
\begin{equation} \label{D.8}
\tilde{A} =  \tilde{A}_R + i \tilde{A}_I,
\end{equation}
and similarly for $\tilde{F}$
\begin{equation} \label{D.9}
\tilde{F} =  \tilde{F}_R + i \tilde{F}_I.
\end{equation}
We find then
\begin{align} \label{D.10}
&(\re{ \tilde{\zeta}})^\trans = \tilde{\xi}^\trans \tilde{A}_R =
\begin{pmatrix} 
\sqrt{\frac{2}{3}} \xi_0 + \xi_3 + \frac{1}{\sqrt{3}} \xi_8, &
\sqrt{\frac{2}{3}} \xi_0 - \xi_3 + \frac{1}{\sqrt{3}} \xi_8, &
\frac{1}{\sqrt{3}}(\sqrt{2} \xi_0 - 2 \xi_8), &
\xi_1, &
\xi_1, &
\xi_4, &
\xi_4, &
\xi_6, &
\xi_6 
\end{pmatrix}\\
 \label{D.11}
&(\im{ \tilde{\zeta}})^\trans = - \tilde{\xi}^\trans \tilde{A}_I =
\begin{pmatrix} 0, & 0, & 0, & \xi_2, & - \xi_2, &  \xi_5, & - \xi_5, & \xi_7, & - \xi_7 \end{pmatrix},  
\end{align}
\begin{equation} \label{D.12}
\begin{split}
&\tilde{F}_R = \tilde{A}^\trans_R \tilde{E} \tilde{A}_R +  \tilde{A}^\trans_I \tilde{E} \tilde{A}_I = \tilde{F}_R^\trans,\\
&\tilde{F}_I = \tilde{A}^\trans_R \tilde{E} \tilde{A}_I -  \tilde{A}^\trans_I \tilde{E} \tilde{A}_R = -\tilde{F}_I^\trans,
\end{split}
\end{equation}

In tables \ref{FR} and \ref{FI} we list the values for$ \tilde{F}_R$ and $ \tilde{F}_I$, respectively.
\begin{table}
\begin{tabular}{c|c||c|c}
$\alpha\;\beta$ & $\tilde{F}_{R \alpha \beta}$ & $\alpha\;\beta$ & $\tilde{F}_{R \alpha \beta}$\\
\hline
0 0 & 
$
\frac{1}{3} \left(3 E_{3 3}+2 \sqrt{3} E_{3 8}+E_{8 8}+2 \sqrt{6} \eta_{3}+2 \sqrt{2} \eta_{8}+2 \eta_{00}\right)$ 
&
3 3 &
$E_{1 1}+E_{2 2}$\\
0 1&
$\frac{1}{3} \left(-3 E_{3 3}+E_{8 8}+2 \sqrt{2} \eta_{8}+2 \eta_{00}\right)$
&
3 4 &
$E_{1 1}-E_{2 2}$\\
0 2 &
$\frac{1}{3} \left(-2 \sqrt{3} E_{3 8}-2 E_{8 8}+\sqrt{6} \eta_{3}-\sqrt{2} \eta_{8}+2 \eta_{00}\right)$
&
3 5 &
$E_{1 4}+E_{2 5}$\\
0 3 & 
$\frac{E_{1 8}+\sqrt{2} \eta_{1}}{\sqrt{3}}+E_{1 3}$
&
3 6 &
$E_{1 4}-E_{2 5}$\\
0 4 &
$\frac{E_{1 8}+\sqrt{2} \eta_{1}}{\sqrt{3}}+E_{1 3}$
&
3 7 &
$E_{1 6}+E_{2 7}$\\
0 5 &
$\frac{E_{4 8}+\sqrt{2} \eta_{4}}{\sqrt{3}}+E_{3 4}$
&
3 8 &
$E_{1 6}-E_{2 7}$\\
0 6 &
$\frac{E_{4 8}+\sqrt{2} \eta_{4}}{\sqrt{3}}+E_{3 4}$
&
4 4 &
$E_{1 1}+E_{2 2}$\\
0 7 &
$\frac{E_{6 8}+\sqrt{2} \eta_{6}}{\sqrt{3}}+E_{3 6}$
&
4 5 &
$E_{1 4}-E_{2 5}$ \\
0 8  &
$\frac{E_{6 8}+\sqrt{2} \eta_{6}}{\sqrt{3}}+E_{3 6}$
&
4 6 &
$E_{1 4}+E_{2 5}$\\
1 1 &
$\frac{1}{3} \left(3 E_{3 3}-2 \sqrt{3} E_{3 8}+E_{8 8}-2 \sqrt{6} \eta_{3}+2 \sqrt{2} \eta_{8}+2 \eta_{00}\right)$
&
4 7 &
$E_{1 6}-E_{2 7}$\\
1 2 &
$\frac{1}{3} \left(2 \sqrt{3} E_{3 8}-2 E_{8 8}-\sqrt{6} \eta_{3}-\sqrt{2} \eta_{8}+2 \eta_{00}\right)$
&
4 8 &
$E_{1 6}+E_{2 7}$\\
1 3 &
$\frac{E_{1 8}+\sqrt{2} \eta_{1}}{\sqrt{3}}-E_{1 3}$
&
5 5 &
$E_{4 4}+E_{5 5}$\\
1 4 &
$\frac{E_{1 8}+\sqrt{2} \eta_{1}}{\sqrt{3}}-E_{1 3}$
&
5 6 &
$E_{4 4}-E_{5 5}$\\
1 5 &
$\frac{E_{4 8}+\sqrt{2} \eta_{4}}{\sqrt{3}}-E_{3 4}$
&
5 7 &
$E_{4 6}+E_{5 7}$\\
1 6 &
$\frac{E_{4 8}+\sqrt{2} \eta_{4}}{\sqrt{3}}-E_{3 4}$
&
5 8 &
$E_{4 6}-E_{5 7}$\\
1 7 &
$\frac{E_{6 8}+\sqrt{2} \eta_{6}}{\sqrt{3}}-E_{3 6}$
&
6 6 &
$E_{4 4}+E_{5 5}$\\
1 8 &
$\frac{E_{6 8}+\sqrt{2} \eta_{6}}{\sqrt{3}}-E_{3 6}$
&
6 7 &
$E_{4 6}-E_{5 7}$\\
2 2 &
$\frac{2}{3} \left(2 E_{8 8}-2 \sqrt{2} \eta_{8}+\eta_{00}\right)$
&
6 8 &
$E_{4 6}+E_{5 7}$\\
2 3 &
$\frac{\sqrt{2} \eta_{1}-2 E_{1 8}}{\sqrt{3}}$
&
7 7 &
$E_{6 6}+E_{7 7}$\\
2 4 &
$\frac{\sqrt{2} \eta_{1}-2 E_{1 8}}{\sqrt{3}}$
&
7 8 &
$E_{6 6}-E_{7 7}$\\
2 5 &
$\frac{\sqrt{2} \eta_{4}-2 E_{4 8}}{\sqrt{3}}$
&
8 8 &
$E_{6 6}+E_{7 7}$\\
2 6 &
$\frac{\sqrt{2} \eta_{4}-2 E_{4 8}}{\sqrt{3}}$\\
2 7 &
$\frac{\sqrt{2} \eta_{6}-2 E_{6 8}}{\sqrt{3}}$\\
2 8 &
$\frac{\sqrt{2} \eta_{6}-2 E_{6 8}}{\sqrt{3}}$\\
\hline
\end{tabular}
\caption{\label{FR} Explicit values of the symmetric matrix $\tilde{F}_{R}$ as defined in \eqref{D.7}.}
\end{table}

\begin{table}
\begin{tabular}{c|c||c|c}
$\alpha\;\beta$ & $\tilde{F}_{I \alpha \beta}$ & $\alpha\;\beta$ & $\tilde{F}_{I \alpha \beta}$\\
\hline
0 1 & 0 & 2 7 & $-\sqrt{\frac{2}{3}} \eta_7 + \frac{2}{\sqrt{3}} E_{78}$\\
0 2 & 0 & 2 8 & $\sqrt{\frac{2}{3}} \eta_7 - \frac{2}{\sqrt{3}} E_{78}$\\
0 3 & $-\sqrt{\frac{2}{3}} \eta_2 - E_{23} - \frac{1}{\sqrt{3}} E_{28}$ & 3 4 & $ 2 E_{12}$\\
0 4 & $\sqrt{\frac{2}{3}} \eta_2 + E_{23} + \frac{1}{\sqrt{3}} E_{28}$ & 3 5 & $-E_{15}+E_{24}$\\
0 5 & $-\sqrt{\frac{2}{3}} \eta_5 - E_{35} - \frac{1}{\sqrt{3}} E_{58}$ & 3 6 & $E_{15}+E_{24}$\\
0 6 & $\sqrt{\frac{2}{3}} \eta_5 + E_{35} + \frac{1}{\sqrt{3}} E_{58}$ & 3 7 & $-E_{17}+E_{26}$\\
0 7 & $-\sqrt{\frac{2}{3}} \eta_7 - E_{37} - \frac{1}{\sqrt{3}} E_{78}$ & 3 8 & $ E_{17}+E_{26}$\\
0 8 & $\sqrt{\frac{2}{3}} \eta_7 + E_{37} + \frac{1}{\sqrt{3}} E_{78}$ & 4 5 & $-E_{15}-E_{24}$\\
1 2 &  0 & 4\; 6 & $E_{15}-E_{24}$\\
1 3 & $-\sqrt{\frac{2}{3}} \eta_2 + E_{23} - \frac{1}{\sqrt{3}} E_{28}$ & 4 7 & $-E_{17}-E_{26}$\\
1 4 & $\sqrt{\frac{2}{3}} \eta_2 - E_{23} + \frac{1}{\sqrt{3}} E_{28}$ & 4 8 & $ E_{17}-E_{26}$\\
1 5 & $-\sqrt{\frac{2}{3}} \eta_5 + E_{35} - \frac{1}{\sqrt{3}} E_{58}$ & 5 6 & $ 2 E_{45}$\\
1 6 & $\sqrt{\frac{2}{3}} \eta_5 - E_{35} + \frac{1}{\sqrt{3}} E_{58}$ & 5 7 & $ -E_{47}+E_{56}$\\
1 7 & $-\sqrt{\frac{2}{3}} \eta_7 + E_{37} - \frac{1}{\sqrt{3}} E_{78}$ & 5 8 & $ E_{47}+E_{56}$\\
1 8 & $\sqrt{\frac{2}{3}} \eta_7 - E_{37} + \frac{1}{\sqrt{3}} E_{78}$ & 6 7 & $ -E_{47}-E_{56}$\\
2 3 & $-\sqrt{\frac{2}{3}} \eta_2 + \frac{2}{\sqrt{3}} E_{28}$ & 6 8 & $ E_{47}-E_{56}$\\
2 4 & $\sqrt{\frac{2}{3}} \eta_2 - \frac{2}{\sqrt{3}} E_{28}$ & 7 8 & $2E_{67}$\\
2 5 & $-\sqrt{\frac{2}{3}} \eta_5 + \frac{2}{\sqrt{3}} E_{58}$\\
2 6 & $\sqrt{\frac{2}{3}} \eta_5 - \frac{2}{\sqrt{3}} E_{58}$\\
\hline
\end{tabular}
\caption{\label{FI} Explicit values of the antisymmetric matrix $\tilde{F}_{I}$ as defined in \eqref{D.7}.}
\end{table}

Suppose now that the potential $V$ allows CP invariance. This holds if and only if there is a basis
where \eqref{7.5} is true for the parameters $\tvec{\xi}$, $\tvec{\eta}$, $E$.
From \eqref{D.11} and table \ref{FI} we see that in the conventional basis \eqref{D.1} this
requires all imaginary parts of the parameters to vanish and vice versa. In this way we recover
the statement, first shown for the THDM in \cite{Gunion:2005ja}, that a potential allows CP
invariance if and only if there is a conventional basis where all parameters are real.


\section{Standard P and CP transformations in the nHDM}
\label{appendixE}

In this appendix we investigate
the standard CP transformation
in the general case of $n \ge 2$ Higgs boson doublets which all 
carry the same hypercharge $y=+1/2$. We denote the
complex doublet fields by
\begin{equation}
\varphi_i(x) = 
\begin{pmatrix} 
\varphi^+_i(x) \\
 \varphi^0_i(x)
 \end{pmatrix}, \qquad i=1,\ldots,n.
\end{equation}

We now introduce the $n \times 2$~matrix of the Higgs-boson fields
\begin{equation} 
\phi = 
\begin{pmatrix} 
\varphi^+_1(x) & \varphi^0_1(x)  \\
\vdots  & \vdots\\
\varphi^+_n(x) & \varphi^0_n(x)
\end{pmatrix} = 
\begin{pmatrix} 
\varphi_1^\trans(x) \\
\vdots \\
\varphi_n^\trans(x) \\
\end{pmatrix} .
\end{equation}
and define the hermitian matrix 
\begin{gather} \label{E.3}
\twomat{K}(x) =  \phi(x) \phi^\dagger(x) =
\begin{pmatrix}
  \varphi_1^{\dagger}(x)\varphi_1(x) & \varphi_2^{\dagger}(x)\varphi_1(x) & \ldots & \varphi_n^{\dagger}(x)\varphi_1(x) \\
  \varphi_1^{\dagger}(x)\varphi_2(x) & \varphi_2^{\dagger}(x)\varphi_2(x) & \ldots & \varphi_n^{\dagger}(x)\varphi_2(x) \\
  \vdots & & \ddots & \ \vdots\\
  \varphi_1^{\dagger}(x)\varphi_n(x) & \varphi_2^{\dagger}(x)\varphi_n(x) & \ldots & \varphi_n^{\dagger}(x)\varphi_n(x)  
\end{pmatrix}.
\end{gather}

A basis for the $n \times n$ matrices is given by the matrices
\begin{equation} \label{A2.4a}
\lambda_\alpha, \quad \alpha = 0,1,\ldots,n^2-1 \;,
\end{equation}
where 
\begin{equation}\label{A2.4b}
\lambda_0 = \sqrt{\frac{2}{n}} \unitmatrix_n
\end{equation}
is the conveniently scaled unit matrix and $\lambda_a$, $a=1,\ldots,n^2-1$, 
are the generalised Gell-Mann matrices. 
An explicit construction and numbering scheme of the generalised Gell-Mann matrices is given
in~\cite{Maniatis:2015gma}.
For making our article self contained we reproduce this numbering scheme here.
We start with defining the $n \times n$ matrix $E_{jk}$ with 1 at the intersection of the $j$th
row and $k$th column and 0 elsewhere. 
In terms of these matrices we construct 
$n^2-1$ hermitian traceless matrices $\lambda_a, a=1,\ldots,n^2-1$, as follows.
With $k=1,\ldots,n-1$ and $j=1,\ldots,k$ we set
\begin{align} 
\label{E5a}
&\lambda_a = E_{j,k+1} + E_{k+1,j}, \quad\text{for } a=k^2+2j-2,\\
\label{E5b}
&\lambda_a = -i E_{j,k+1} + i E_{k+1,j}, \quad\text{for } a=k^2+2j-1.\\
\intertext{In addition we construct $n-1$ diagonal matrices}
\label{E5c}
&\lambda_{(l+1)^2-1} = \sqrt{\frac{2}{l(l+1)}}
\left[ \left( \sum_{j=1}^{l}  E_{jj} \right) - l\cdot E_{l+1\, l+1} \right],\\ 
&\qquad 1 \le l \le n-1. \nonumber
\end{align}
An easy way to remember this numbering scheme
is as follows. We draw an $n \times n$ rectangular
lattice and insert the numbers $\alpha = 0,1,\ldots,n^2-1$
as shown in Fig.~\ref{fig-numbering}.
If~$\alpha$ is the upper (lower) number in an off-diagonal
rectangle then $\lambda_\alpha$ gets a 1 ($-i$) in this place,
1 ($+i$) in the transposed place, and zero elsewhere.
If $\alpha$ is in a diagonal rectangle $\lambda_\alpha$ is given 
by \eqref{E5c} for $\alpha>0$ and
by \eqref{A2.4b} for $\alpha = 0$.
\begin{figure}[t!]
\begin{small}
\begin{tabular}{| >{\centering}m{36pt} | >{\centering}m{36pt} | >{\centering}m{36pt} | >{\centering}m{36pt} | >{\centering}m{36pt} | >{\centering}m{37pt} |}
\hline
0 & $\begin{matrix} 1\\ 2\end{matrix}$ & $\begin{matrix}4\\ 5\end{matrix}$ & $\begin{matrix}9\\ 10 \end{matrix}$ & $\cdots$ & $\begin{matrix}(n\!-\!1)^2 \phantom{\!{\tiny +}\!0}\\ (n\!-\!1)^2\!{\tiny +}\!1 \end{matrix}$\\
\tabularnewline
\hline
 & 3  & $\begin{matrix}6\\ 7\end{matrix}$ & $\begin{matrix}11\\ 12\end{matrix}$ & $\cdots$ & $\begin{matrix}(n\!-\!1)^2\!{\tiny +}\!2\\ (n\!-\!1)^2\!{\tiny +}\!3 \end{matrix}$ \\
\tabularnewline
\hline 
 &  & 8  & $\begin{matrix}13\\ 14\end{matrix}$ & $\cdots$ & $\small{\begin{matrix}(n\!-\!1)^2\!{\tiny +}\!4\\ (n\!-\!1)^2\!{\tiny +}\!5\end{matrix}}$\\
\tabularnewline
\hline 
 &  &  & 15  &  $\cdots$ & $\begin{matrix}(n\!-\!1)^2\!{\tiny +}\!6\\ (n\!-\!1)^2\!{\tiny +}\!7\end{matrix}$\\
\tabularnewline
\hline
$\vdots$ & $\vdots$ & $\vdots$ & $\vdots$ & $\ddots$ & $\vdots$ \\
\tabularnewline
\hline 
 & & & & $\cdots$ & $\begin{matrix}  \\ n^2-1 \end{matrix}$\\
\tabularnewline
\hline 
\end{tabular}
\end{small}
\caption{\label{fig-numbering}Numbering scheme for the generalised Gell-Mann matrices
$\lambda_\alpha$ ($\alpha=0, \ldots, n^2-1$).}
\end{figure}
We have
\begin{equation} \label{A2.4c}
\tr (\lambda_\alpha \lambda_\beta) = 2 \delta_{\alpha \beta},\qquad
\tr (\lambda_\alpha) = \sqrt{2 n}\; \delta_{\alpha 0}.
\end{equation}

The matrix~$\twomat{K}$~\eqref{E.3} can be written in the basis of the scaled unit matrix
and the generalised Gell-Mann matrices as 
\begin{equation} \label{A2.5}
\twomat{K}(x) = \frac{1}{2} K_\alpha(x) \lambda_\alpha ,
\end{equation}
where the real coefficients $K_\alpha$ are given by 
\begin{equation} \label{A2.6}
K_\alpha(x) = K_\alpha^*(x) = \tr (\twomat{K}(x) \lambda_\alpha).
\end{equation}

The standard parity transformation, $\Ps$, reads
\begin{equation}
\varphi_i(x) \stackrel{\Ps}{\longrightarrow} \varphi_i(x'),\qquad i=1,\ldots,n,
\end{equation}
with $x$ and $x'$ given in \eqref{4.101}. 
For the bilinears we get from \eqref{E.3} and \eqref{A2.6}
\begin{equation}
K_\alpha(x) \stackrel{\Ps}{\longrightarrow} K_\alpha(x').
\end{equation}

Next we consider the standard CP transformation
\begin{equation} \label{ACPsfields}
\varphi_i(x) \stackrel{\CPs}{\longrightarrow} \varphi_i^*(x'), \qquad i=1,\ldots n.
\end{equation}
This corresponds in terms of the matrices $\phi$ and $\twomat{K}$ to 
\begin{equation} \label{C.10A}
\begin{split}
&\phi(x) \stackrel{\CPs}{\longrightarrow} \phi'(x) =  \phi^*(x'), \\
&\twomat{K}(x)   \stackrel{\CPs}{\longrightarrow} \twomat{K}'(x) = \twomat{K}^*(x') = \twomat{K}^\trans(x').
\end{split}
\end{equation}
With this we see that the bilinears transform under $\CPs$ as
\begin{equation} \label{ACPsK}
\begin{split}
&K_0(x) \stackrel{\CPs}{\longrightarrow} K'_0(x) = \tr (\twomat{K}'(x) \lambda_0) = \tr (\twomat{K}^\trans(x') \lambda_0) =  K_0(x') ,\\ 
&K_a(x)  = \tr(\twomat{K}(x) \lambda_a)  \stackrel{\CPs}{\longrightarrow} K'_a(x) = \tr (\twomat{K}'(x) \lambda_a)= \tr (\twomat{K}^\trans(x') \lambda_a) 
= \tr (\twomat{K}(x') \lambda_a^\trans)
= C_{ab}^s K_b(x') ,
\end{split}
\end{equation}
where we define the $(n^2-1) \times (n^2-1)$ matrix $\hat{C}^s$ by 
\begin{equation} \label{ARdefCPs}
\lambda_a^\trans = \hat{C}_{ab}^s \lambda_b, \qquad \hat{C}^s = \left( \hat{C}_{ab}^s \right),
\qquad a,b \in \{1,\ldots,n^2-1\}.
\end{equation}

Let us study the $(n^2-1) \times (n^2-1)$ matrices $\hat{C}^s$ in detail. 
For given $n$  there are
$n^2-1$ generalised Gell-Mann matrices $\lambda_b$; $b=1,\ldots,n^2-1$. 
These generalised Gell--Mann matrices are either 
symmetric or antisymmetric matrices. Hence,
the matrix $\hat{C}^s$ is diagonal and 
for every symmetric matrix $\lambda_a$ we
get a diagonal entry $+1$ in $\hat{C}^s$ and
for every antisymmetric matrix $\lambda_a$ we have
a diagonal entry $-1$. The general form which we find for the $(n^2-1)\times (n^2-1)$ matrix $\hat{C}^s$ is therefore
\begin{equation}
\hat{C}^s = \diag(\pm 1, \pm 1, \ldots, \pm 1).
\end{equation}
Note that this form of the matrix $\hat{C}^s$ ensures that
two subsequent standard $\CPs$ transformations give back the original 
bilinears:
\begin{equation}
K_\alpha(x)  \stackrel{\CPs \circ \CPs}{\longrightarrow} K_\alpha(x).
\end{equation}
Of course, we see this also immediately at the level of the fields from \eqref{ACPsfields}
\begin{equation} \label{E15a}
\varphi_i(x) \stackrel{\CPs}{\longrightarrow} \varphi_i^*(x') \stackrel{\CPs}{\longrightarrow} \varphi_i(x) .
\end{equation}

Now let us count the number of antisymmetric generalised Gell-Mann matrices. In the
representation as given in Fig.~\ref{fig-numbering} the antisymmetric matrices are those of \eqref{E5b}.
For a given $n$ the second row of table~\ref{lambdaanti} gives 
the indices $a$ of the 
antisymmetric matrices $\lambda_a$. 
That is, for given $n$ all matrices with indices listed in the second row up to the entry $n$ are the antisymmetric ones.
The third row of table~\ref{lambdaanti} shows the total number of antisymmetric matrices
for a given $n$.
\begin{table}
\begin{tabular}{cccccccc}
\hline
$n$ & 2 & 3 & 4& 5 & 6 & $\ldots$ & n\\
\hline
$a$ & 2,\; & 5,7,\; & 10,12,14\; & 17,19,21,23,\; & 26,28,30,32,34, & $\ldots$&,
\begin{small}$n^2\!-\!2n\!+\!2, n^2\!-\!2n\!+\!4, \ldots, n^2\!-\!2$\end{small}\\
\hline
\# antisymmetric & 1 & 3 & 6 & 10 & 15 & $\ldots$ & $n\cdot (n-1)/2$\\
\hline
\end{tabular}
\caption{\label{lambdaanti} Counting of the number of antisymmetric generalised Gell-Mann matrices $\lambda_a$ as function of $n$. 
The second row gives the indices of the antisymmetric matrices in the numbering scheme shown in Fig. \ref{fig-numbering}. For given $n$
all indices listed in the second row up to $n$ correspond to the antisymmetric generalised Gell-Mann matrices for this $n$.
The third row gives the
total number of antisymmetric matrices depending on $n$.
For instance for $n=4$ we get the antisymmetric matrices $\lambda_a$ with
$a= 2, 5, 7, 10, 12, 14$, that is, in total $6$ antisymmetric matrices.}
\end{table}

Let us look at the simplest cases. 
For  $n=2$, that is, for the THDM, there is only one antisymmetric matrix $\lambda_2$$(a=2)$, which is
the second Pauli matrix. The total number of antisymmetric matrices is one. 
Therefore we get from \eqref{ARdefCPs} in this case
for the $\CPs$ transformation matrix, as shown in section 3 of \cite{Maniatis:2007vn},
\begin{align}
&\text{THDM:} \qquad &&\hat{C}^s = \diag(1, -1, 1 ). \label{CPsTHDM}\\
\intertext{\noindent
This is a reflection in the space of the bilinears $K_a$
since $\det(\hat{C}^s)=-1$.
For $n=3$, relevant for the 3HDM, the Gell--Mann
matrices $\lambda_a$ with $a=2, 5, 7$ are antisymmetric and all remaining matrices 
symmetric, as listed in table \ref{lambdaanti}.
Hence, the matrix $\hat{C}^s$ has the explicit form, as already given in \eqref{CPs3HDM},
} 
&\text{3HDM:} \qquad &&\hat{C}^s = \diag ( 1, -1, 1,1,-1,1,-1,1).\\
\intertext{\noindent
This is again a reflection in $K$ space, that is, $\det (\hat{C}^s) = -1$.
Proceeding with  the 4HDM,
following table \ref{lambdaanti}, we see that the corresponding 
$\CPs$-transformation matrix is
}
&\text{4HDM:} \qquad && \hat{C}^s = \diag ( 1, -1, 1,1,-1,1,-1,1,1,-1,1,-1,1,-1,1). \\
\intertext{\noindent
Here we  obviously have $\det(\hat{C}^s)= +1$ unlike the cases
of the THDM and the 3HDM where the determinant is $-1$. 
For the general case of $n$ let $I_{\text{as}}^n$ be the set of the indices of the antisymmetric generalised
Gell--Mann matrices $\lambda_a$ as given in table \ref{lambdaanti},
\begin{equation}
I_{\text{as}}^n = \{ k^2-2k+2,\;k^2-2k+4,\;\ldots,\;k^2-2 \; | \; k=2,\ldots,n \},
\end{equation}
that is,
\begin{equation}
I_{\text{as}}^n = \{ 2,\; 5,7,\; 10,12,14,\; 17,19,21,23,\; \ldots, n^2-2 \} .
\end{equation}
$I_{\text{as}}^n$ has $n(n-1)/2$ elements. The matrix of the standard $\CPs$ transformation reads
$\hat{C}^s = \left( \hat{C}_{ab}^s \right)$ with
}\\
&\text{nHDM:} \qquad && \hat{C}_{ab}^s= 
\begin{cases}
-1 & \text{ for } a=b \in I_{\text{as}}^n, \\
+1 & \text{ for } a=b\notin I_{\text{as}}^n,\\
\phantom{+}0 & \text{ for } a\neq b,
\end{cases} \label{Rsbar}
\end{align}
where $a,b \in \{1, \ldots, n^2-1\}$.\\

To conclude, we find that the standard $\CPs$ transformation of the nHDM \eqref{ACPsfields}
corresponds for the bilinears $K_0(x)$, $K_a(x)$ to a linear transformation; see
\eqref{C.10A}, \eqref{ACPsK}, \eqref{Rsbar}. The transformation of the $K_a(x)$ is governed 
by an $(n^2-1) \times (n^2-1)$ diagonal matrix $\hat{C}^s$. 
The diagonal elements of $\hat{C}^s$ are $\pm 1$. The indices of the $-1$ elements
are obtained from the second row of table \ref{lambdaanti} and 
correspond to the antisymmetric generalised Gell-Mann matrices. The total number
of diagonal elements in $\hat{C}^s$ equal to $-1$ is $n(n-1)/2$. 


\end{document}